%% file: paper.tex
\documentclass{mytemplate}
\title{Time-Varying Causal Survival Learning under Staggered Adoption} 
\author{%
  Xiang Meng\thanks{Department of Data Science, Dana-Farber Cancer Institute;
  Department of Biostatistics, Harvard T.H. Chan School of Public Health.}
  \and
  Iavor Bojinov\thanks{Technology and Operations Management Unit,
  Harvard Business School.}%
}

\begin{document}
\maketitle
\begin{abstract}

Staggered adoptions are settings where study objects convert from control to treatment at different times. 
Often the outcome is a time to an event, such as survival after an organ transplant or the time until an employee becomes a regular user of a new AI tool. Although many papers have studied staggered adoptions on repeatedly measured outcomes, little work has addressed survival outcomes, which are observed at most once and may be censored. We give causal assumptions under which a time-dependent survival model identifies a heterogeneous treatment effect, which tells who benefits most. Because a direct Cox learner is biased when baseline risk is learned flexibly, we introduce a time-varying, event-risk propensity score.  Constructing this propensity score is nontrivial because both treatment and outcome are times to events, so it must track which individuals have not yet experienced the event and the treatment status of each at the time. Unlike existing propensity scores, centering by this propensity score makes the Cox partial-likelihood equation Neyman orthogonal. We establish the double machine learning property and the corresponding convergence rate, and an influence-function-based variance estimator.  Simulations show that orthogonalization reduces error and supports near-nominal inference when the baseline risk is complex and the samples are large, whereas direct Cox estimation is more precise when baseline risk is linear. We illustrate the framework using lawyer-level Copilot adoption data and the Stanford Heart Transplant study.
\end{abstract}

\section{Introduction}
Many treatments and interventions are adopted at different times \citep{athey2022design}. A patient waits for an organ transplant, a customer starts a free trial, or a professional begins using a new software tool \citep{brynjolfsson2025generative}. Each unit therefore contributes untreated time before adoption and treated time afterward. When the outcome is itself a time to event, assigning everyone who eventually adopts to a baseline treated group incorrectly labels pre-adoption person-time as treated and can bias the treatment comparison.  This immortal-time bias is well documented in epidemiology \citep{suissa2008immortal}. In the Stanford Heart Transplant programme, the classical medical example \citep{crowley1977covariance, zhu2021stanford}, patients with severe heart disease received a transplant only when a suitable donor became available, after waits ranging from days to more than a year \citep{evans1986donor, o1992cardiac}. Early comparisons suggested a large survival benefit of transplantation, whereas reanalyses that counted waiting time as untreated found much weaker evidence \citep{mantel1974evaluation, turnbull1974survivorship}. Decisions such as how to allocate organs, or whether a firm should invest in rolling out a tool, therefore depend on methods that respect adoption timing.

Staggered-adoption designs compare treated units with units that have not yet adopted \citep{athey2022design}. Existing designs are largely developed for repeatedly measured outcomes, such as annual GDP, for which the same unit contributes several pre- and post-adoption observations. A survival outcome is different: the event time is observed once, or is censored. The time-dependent Cox model handles adoption timing by updating treatment status at each event time \citep{fisher_time-dependent_1999}, and heterogeneous-effect learners for survival outcomes allow flexible covariate adjustment \citep{cui2023estimating, gao2025estimating}. However, valid inference from the former requires a correctly specified low-dimensional model for baseline risk, and the latter assume that treatment is fixed at baseline. What is missing is a method that handles staggered adoption, learns baseline risk flexibly, and still gives valid confidence intervals for heterogeneous effects.

Similar timing arises outside medicine. In our main application, we study whether a lawyer's first use of an internally developed AI assistant changes the time to becoming a regular user of a second AI tool, Microsoft Copilot, and whether this differs between junior and senior lawyers. As in the transplant study, there is no common date that separates treated and untreated units. Section~\ref{sec:copilot-application} shows that ignoring adoption timing inflates the estimated effect for both groups and hides a junior--senior difference that the time-varying analysis detects. We analyse the Copilot application in the main text and the Stanford Heart Transplant study in the Appendix.

We connect the two settings by expressing the staggered-adoption restrictions in terms of counterfactual hazards. No anticipation makes the pre-adoption hazard equal to the never-treated hazard, and invariance to treatment history makes the post-adoption hazard depend on current treatment status rather than the exact adoption date. Together with standard causal assumptions and the proportional-hazards restriction, these conditions give a causal interpretation to a conditional log hazard ratio. This interpretation requires care because hazard ratios compare subjects who remain at risk over time \citep{hernan2010hazards, aalen2015cox, martinussen2020hazard}, as discussed in Section~\ref{sec:causal-target}.

Flexible estimation of baseline risk creates a second problem. In a direct time-dependent Cox fit, the estimation error from a regularized or flexible baseline-risk model passes directly into treatment-effect estimate. Because such models typically converge more slowly than $n^{-1/2}$, the treatment-effect estimate also converges too slowly for confidence intervals to attain nominal coverage. We therefore use double machine learning (DML) to construct a partial-likelihood score whose first derivative with respect to the nuisance functions is zero \citep{chernozhukov2018double}. The key is to find the correct propensity score for treatment centering. We show that this is the probability of having adopted treatment among subjects who experience the event at time $t$, conditional on their covariates. This differs from the usual adoption propensity because it also accounts for survival selection and the relative event hazards. Centering by this event propensity makes the Cox partial-likelihood score Neyman orthogonal.

The paper makes three contributions. First, it formulates a staggered-adoption design for time-to-event outcomes and states the assumptions under which the heterogeneous log hazard ratio has a causal interpretation. Second, and most importantly, it derives the event-risk propensity that makes the Cox partial-likelihood score Neyman orthogonal. The resulting estimator has a second-order nuisance remainder and a DML convergence rate for a finite-dimensional heterogeneous effect. The event-time-pooled modification of \cite{gao2025estimating} does not generally satisfy the required risk-set centering identity and therefore need not orthogonalize the Cox partial-likelihood score. Third, we establish asymptotic linearity and provide conditions for valid sandwich variance estimation and pointwise confidence intervals for the heterogeneous log hazard ratio.

Our simulations show when orthogonalization is useful. When baseline log risk is linear, an ordinary time-dependent Cox regression is simpler and more precise, because a correctly specified low-dimensional model leaves no first-order bias to remove. When baseline risk is non-linear and fitted flexibly, TV-CSL performs better at sufficiently large sample sizes. At $n=2000$, its MSE is less than half that of Cox, and its coverage stays near nominal while that of Cox falls well below. An additional simulation shows that the event-time propensity reduces bias and improves coverage relative to marginal centering as the treatment effect increases.

In the Copilot application, we find that first use of the firm's AI assistant is associated with a higher rate of becoming a regular Copilot user. Using TV-CSL with the time-varying propensity, the estimated hazard ratio is $1.51$ (95\% CI: $1.28$--$1.79$) for senior lawyers and $1.23$ ($1.02$--$1.48$) for junior lawyers. These findings suggest a positive association between adoption of one AI tool and subsequent regular use of another.  The causal interpretation of these estimates relies on the unconfoundedness, the proportional-hazards, and event-time overlap assumptions, whose limitations we discuss below.

\subsection{Literature Review}

Our work builds on survival analysis, staggered-adoption designs, and heterogeneous treatment-effect estimation. Survival analysis provides tools for time-to-event data, while recent causal work studies treatments that change over time. Their intersection, causal inference for survival outcomes with a one-time treatment adopted at a stochastic time, is less developed.

Several strands of causal inference address survival outcomes. \cite{robins1992aft} studies time-dependent accelerated failure-time models in the presence of confounding, while \cite{abbring2003nonparametric} study identification of treatment effects in duration models. \cite{hernan2010hazards}, \cite{aalen2015cox}, and \cite{martinussen2020hazard} discuss the causal interpretation of hazard ratios, and \cite{vansteelandt2024assumption} develop assumption-lean Cox regression. \cite{gran2010sequential} and \cite{keogh2023sequential} study sequential Cox models and trial-based approaches for handling time-dependent confounding. Our treatment instead changes once, at a stochastic time, and then remains on. Each failure therefore compares already-treated and not-yet-treated subjects in the contemporaneous risk set.

Treatments with random timing have been studied extensively in the difference-in-differences literature. \cite{athey2022design} and \cite{goodman2021difference} emphasize the role of treatment timing in panel designs, while \cite{sun2021estimating} studies bias from effect heterogeneity across adoption cohorts and \cite{imai2021use} develops related methods for panel data. These methods do not directly address a once-observed, possibly censored event time.

For Cox regression, \cite{andersen1982cox} establish the large-sample theory using counting processes, and \cite{fisher_time-dependent_1999} discuss time-dependent covariates. \cite{fang2017testing} develop inference using decorrelated scores for high-dimensional Cox models. Other causal uses of the Cox model answer different questions. \cite{shaikh2021randomization} develop randomization tests for staggered adoption, whereas we estimate effect heterogeneity. \cite{zhu2022causal} consider electronic-health-record settings with longitudinal covariates; our identifying assumptions instead concern one-time adoption and baseline confounding.

For heterogeneous treatment-effect estimation with time-to-event outcomes, \cite{cui2023estimating} develop causal survival forests using doubly robust estimating equations, and \cite{xu_estimating_2024} propose censoring-unbiased transformations that can be combined with existing HTE learners. \cite{xu_treatment_2022} review extensions of metalearners to right-censored outcomes. \cite{gao2025estimating} develop a DML construction for exponential families and discuss the Cox model as a special case. Unlike causal survival forests formulated for baseline treatment and the event-pooled propensity modification of \cite{gao2025estimating}, we study treatment adopted at a stochastic time and derive the event-time propensity required for Neyman orthogonality of the Cox partial likelihood.

Section 2 introduces the causal model and target. Section 3 reviews the time-dependent Cox partial likelihood, and Section 4 develops TV-CSL and its inference theory. Section 5 reports the simulations, Section 6 describes the Copilot adoption application, and Section 7 concludes. The event-propensity derivation, proofs, complete simulation tables, and Stanford Heart Transplant analysis are in the Appendix.

\section{Problem Setup}
\subsection{Notation and observed data}

Let capital letters denote random variables and lowercase letters their realizations. For subject $i$, let $T_i(a)\in[0,\infty]$ be the potential event time if treatment is first adopted at time $a\in[0,\infty]$; $a=\infty$ denotes no adoption during follow-up. We observe the adoption time $A_i$ and baseline covariates $X_i\in\mathbb R^p$. The associated predictable treatment process is
\[
W_i(t)=\mathbf 1(A_i<t).
\]
The strict inequality records treatment status immediately before $t$ and is immaterial when time is continuous and ties have probability zero. In the application of Section~\ref{sec:copilot-application}, a unit is a lawyer, $X_i$ contains baseline characteristics such as rank, office, and practice group, $A_i$ is the week of first use of the firm's AI assistant, and $T_i$ is the week in which the lawyer becomes a regular Copilot user.

Let $C_i$ be the censoring time, let $U_i=T_i\wedge C_i$ be the observed follow-up time, and let $\Delta_i=\mathbf 1(T_i\leq C_i)$ indicate an observed event. We observe $O_i=(X_i,A_i,U_i,\Delta_i).$
Censoring can arise from loss to follow-up or administrative study end; for a censored subject, we know only that $T_i>C_i$. We define the at-risk and event counting processes as
\[
Y_i(t)=\mathbf 1(t\leq U_i),\qquad
N_i(t)=\mathbf 1(U_i\leq t,\Delta_i=1).
\]

We adapt three standard causal assumptions \citep{rubin1974estimating}.

\begin{assumption}[SUTVA]
\label{ass:sutva}
(i) No interference: a subject's potential event times do not depend on other subjects' adoption times, and treatment has a single well-defined version. (ii) Consistency: the observed event time satisfies $T_i=T_i(A_i)$.
\end{assumption}
SUTVA rules out spillovers between subjects and multiple versions of treatment; we need it so that each subject has a single potential event time for each adoption date. In the application, it requires that a colleague's use of the AI assistant does not change a lawyer's own Copilot uptake.

\begin{assumption}[Unconfounded adoption time]
\label{ass:unconfounded}
For every $a$,
\[
A_i\indep T_i(a)\mid X_i.
\]
\end{assumption}
Thus the measured baseline covariates are sufficient to compare subjects with different adoption times. We interpret $A_i$ as a latent adoption time, the time at which subject $i$ would adopt if still event-free and under observation, so it is defined even for subjects who fail first. When adoption cannot follow the event, as for death, $A_i$ is observed only if $A_i<U_i$; all estimating equations below use $A_i$ only through $W_i(t)$ for $t\leq U_i$.

\begin{assumption}[Overlap]
\label{ass:overlap}
There exist $\epsilon>0$ and
$0<t_0<\bar t<\infty$ such that, for almost
every $x$ and all $t\in[t_0,\bar t]$,
\[
\epsilon\leq
\pi_t(x):=P(A<t\mid X=x)
\leq1-\epsilon,
\]
and
\[
P(U\geq\bar t\mid X=x,A)\geq\epsilon
\]
for almost every adoption time $A$.
\end{assumption}

The first requirement ensures variation in
treatment adoption. The second ensures that
both treatment groups can remain represented
in the risk sets throughout the analysis
horizon; without it, survival or censoring
could eliminate treatment possibility of one arm from the comparisons
underlying the Cox partial likelihood.

\subsection{Hazard restrictions and causal target}
\label{sec:causal-target}

Without parametric restrictions, write
\[
X_i\sim f_X,\qquad A_i\mid X_i\sim k(\cdot\mid X_i),
\qquad T_i(a)\mid X_i\sim f(\cdot\mid a,X_i).
\]
Here $f_X$ describes the population of units, $k$ the adoption-time mechanism, and $f$ the distribution of the potential event time under adoption at $a$.
For a fixed adoption time $a$ and covariate value $x$, the counterfactual hazard is
\[
h(t\mid a,x)=\lim_{\delta\downarrow0}
\frac{P\{t\leq T_i(a)<t+\delta\mid T_i(a)\geq t,X_i=x\}}{\delta}.
\] 
It captures the instantaneous event rate at time $t$ among subjects who would still be event-free at $t$ had they adopted at $a$. We work on this hazard scale because it accommodates censoring and because the Cox partial likelihood compares exactly these rates within each risk set. The hazard and event-time distribution
uniquely determine one another. For a
generic hazard $h(t)$, let
$\Lambda(t)=\int_0^t h(s)\,ds$ denote the cumulative hazard. Then
\[
1-F(t)=e^{-\Lambda(t)},
\qquad
f(t)=h(t)e^{-\Lambda(t)},
\]
and conversely
$h(t)=f(t)/\{1-F(t)\}$
\citep{klein_survival_1997}.

There is one counterfactual hazard for every possible adoption time. Two restrictions reduce this collection to hazards under current treatment status.

\begin{assumption}[No anticipation]
\label{ass:no-anticipation}
For $t<a$,
\[
h(t\mid a,x)=h(t\mid\infty,x).
\]
\end{assumption}
No anticipation means before adoption, a future adoption date does not change the current hazard \citep{abbring2008event}. It rules out changes in event risk made in advance of a planned adoption date, and it is what allows pre-adoption person-time to be counted as untreated. It is plausible when adoption dates are not known in advance, as when a transplant depends on donor availability; it would fail if, for example, lawyers who planned to start using the assistant began experimenting with Copilot beforehand.

\begin{assumption}[Invariance to adoption history]
\label{ass:history-invariance}
For $t>a$,
\[
h(t\mid a,x)=h(t\mid0,x).
\]
\end{assumption}
This assumption means after adoption, the hazard depends on current treatment status but not on the exact adoption date or time since adoption. Related restrictions are used in staggered-adoption designs; see Section 3.2 of \cite{athey2022design}. It rules out effects that build up or fade with time since adoption, such as learning, or elevated surgical risk shortly after transplantation, and it is what reduces the many adoption-specific hazards to a single treated hazard. In the Copilot application, it requires that the effect of having used the assistant not grow or decay over the following weeks.

Define the untreated and treated hazards by
\[
h_0(t\mid x)=h(t\mid\infty,x),\qquad
h_1(t\mid x)=h(t\mid0,x).
\]
Under Assumptions 4 and 5,
\[
h(t\mid a,x)=h_0(t\mid x)
\left\{\frac{h_1(t\mid x)}{h_0(t\mid x)}\right\}^{\mathbf 1(a<t)}.
\]
We assume the proportional-hazards model \citep{cox1972regression}
\begin{equation}
\begin{split}
h_0(t\mid x)&=\lambda_0(t)e^{\eta_0(x)},\\
h_1(t\mid x)&=\lambda_0(t)e^{\eta_1(x)}.
\end{split}
\end{equation}
Here $\lambda_0(t)$ is the unspecified
baseline hazard,
$\Lambda_0(t)=\int_0^t\lambda_0(s)\,ds$
is its cumulative hazard, and $\eta_w(x)$
is the covariate-dependent log relative
risk under treatment status $w$.
The baseline hazard cancels from the
treatment contrast
\begin{equation}
\tau_0(x)
=\log\left\{\frac{h_1(t\mid x)}{h_0(t\mid x)}\right\}
=\eta_1(x)-\eta_0(x).
\end{equation}
We call $\tau_0(x)$ the heterogeneous log hazard ratio (HLHR), following the use of covariate-specific hazard-ratio targets in survival HTE estimation \citep{kalbfleisch2011statistical, gao2025estimating}. We use the proportional-hazards model as a natural starting point because its partial likelihood removes $\lambda_0$ so no model for how risk changes over time is needed. Extending the approach to effect measures that do not require proportional hazards, such as the restricted mean survival time, is left for future work.

We assume that the HLHR is linear in a fixed, user-chosen basis $p(x)\in\mathbb R^d$:
\begin{equation}
\label{eq:parametric-HLHR}
\tau_0(x)=p(x)^\top\beta_0.
\end{equation}
The basis may contain original covariates, transformations, or prespecified interactions. This restriction keeps the target finite-dimensional while allowing $\eta_0(x)$ to be learned flexibly.  We fix the basis in advance for two reasons. First, the scientific question usually concerns a few prespecified effect modifiers, such as seniority in our application, and $\beta_0$ answers it directly. Second, a finite-dimensional target permits root-$n$ inference even when $\eta_0$ is learned flexibly. The cost is that heterogeneity outside the span of $p(x)$ is averaged over rather than detected. We assume that $p(x)$ contains an intercept, as in our simulations and application.

The causal interpretation of a hazard ratio requires care because conditioning on survival can induce selection over time \citep{hernan2010hazards}. Under the proportional-hazards model above, this selection is incorporated into the model through the covariate-specific, time-invariant contrast \(\tau_0(x)\); when proportional hazards fails, however, \(\tau_0(x)\) should instead be viewed as a working summary rather than a time-invariant causal effect.

The observed time-varying hazard can now be written as
\begin{equation}
\label{eq:ordinary-cox-hazard}
h(t\mid A_i=a,X_i=x)
=\lambda_0(t)\exp\{\eta_0(x)+\mathbf 1(a<t)p(x)^\top\beta_0\}.
\end{equation}

Our parameter of interest is $\beta_0$. $\eta_0$ is the nuisance parameter. Section~\ref{sec:s-lasso-review} first reviews estimation procedure of $\beta$ under the time-varying hazard when $\eta$ is simple. Section~\ref{sec:DML} then develops a valid inference procedure for $\beta_0$ when $\eta_0$ has a complex structure where  an orthogonal score equation is derived for double machine learning.

\section{Time-Dependent Cox Partial Likelihood}
\label{sec:s-lasso-review}

This section reviews standard results~\citep{fisher_time-dependent_1999, kalbfleisch2011statistical}. Its purpose is to show what changes under staggered adoption and is to isolate the problem that motivates Section~\ref{sec:DML}. In short, every subject in each risk set must enter with the treatment status held just before that event time, $W_j(U_i)$, rather than a status fixed at baseline.    

With treatment fixed at baseline, consider
\begin{equation}
\label{eq:ordinary-cox-hazard-fixed}
h(t\mid w,x)=\lambda_0(t)\exp\{\eta_0(x)+w\tau_0(x)\},
\end{equation}
with the same treatment status used throughout follow-up. 
At an observed event time $U_i$, define the risk set
\begin{equation}
\label{eq:risk-set}
\mathcal R_i
=\{j:Y_j(U_i)=1\}
=\{j:U_j\geq U_i\},
\end{equation}
consisting of subjects who remain event-free and uncensored immediately before $U_i$.

Conditional on one subject in $\mathcal R_i$ failing (meaning an event is observed) at $U_i$, the partial-likelihood contribution is the hazard of subject $i$ divided by the sum of the hazards over $\mathcal R_i$. The log of the product of these contributions is \citep{cox1972regression, andersen_statistical_1993}
\begin{equation}
\label{eq:pl-original-data}
\begin{split}
\operatorname{pl}_n(\tau,\eta_0)
=\frac1n\sum_{\Delta_i=1}\Bigg[
&\eta_0(X_i)+W_i\tau(X_i)\\
&-\log\left\{\sum_{j\in\mathcal R_i}
\exp\bigl(\eta_0(X_j)+W_j\tau(X_j)\bigr)\right\}
\Bigg].
\end{split}
\end{equation}

For staggered adoption, treatment status must instead be read at each event time. The key distinction is the treatment
indicator: the fixed model uses the same
$W_j$ at every event time, whereas
staggered adoption requires the
time-updated $W_j(U_i)$ for each subject
in the risk set. The resulting partial
likelihood is
\begin{equation}
\label{eq:pl-tv-original-data}
\begin{split}
\operatorname{pl}_n(\tau,\eta_0)
=\frac1n\sum_{\Delta_i=1}\Bigg[
&\eta_0(X_i)+W_i(U_i)\tau(X_i)\\
&-\log\left\{\sum_{j\in\mathcal R_i}
\exp\bigl(\eta_0(X_j)+W_j(U_i)\tau(X_j)\bigr)\right\}
\Bigg].
\end{split}
\end{equation}

\begin{figure}[t]
\centering
\includegraphics[width=\textwidth]{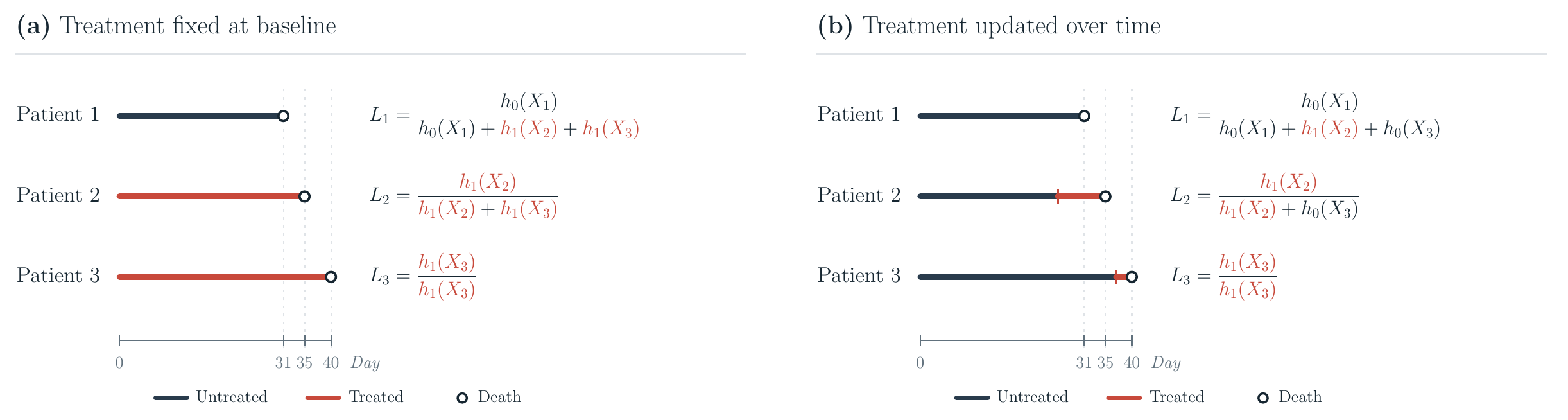}
\caption{Partial-likelihood contributions with fixed and time-varying treatment. Black denotes untreated person-time and red denotes treated person-time. Panel (a) treats eventual adopters as treated from baseline, which shows that the control group appears to have shorter observed survival times on average. In contrast, panel (b) updates treatment status at the actual adoption times, which shows that treated time before death is shorter than untreated time on average. This contrast illustrates how ignoring treatment timing changes the risk-set comparisons and why each Cox partial-likelihood denominator must use the treatment status of every subject at the corresponding event time.}
\label{fig:pl-comparison}
\end{figure}

In Equation~\ref{eq:pl-original-data}, each subject's contribution is fixed at either $\eta_0(X_j)$ or $\eta_1(X_j)$ across all event times. In contrast, Equation~\ref{eq:pl-tv-original-data} allows the same subject's contribution to change from $\eta_0(X_j)$ before adoption to $\eta_1(X_j)$ afterward. Figure~\ref{fig:pl-comparison} illustrates the resulting change in the risk-set denominators.




We impose the following standard
regularity conditions for time-dependent
Cox regression.

\begin{condition}[Cox regularity]
\label{cond:cox}
\begin{enumerate}[(i)]

\item The observations are i.i.d. by subject.
Treatment status $W_i(t)$ is determined
immediately before $t$, and the observed
event process has conditional intensity
\[
Y_i(t)e^{\eta_{W_i(t)}(X_i)}
\,d\Lambda_0(t).
\]

\item The cumulative baseline hazard
$\Lambda_0$ is continuous on $[0,\bar t]$ and $\Lambda_0(\bar t)$ is finite.

\item The population hazard-weighted
risk-set denominator
\[
c_t:=
E[Y(t)e^{\eta_{W(t)}(X)}]
\]
satisfies
$\inf_{t\in[0,\bar t]}c_t>0$.

\end{enumerate}
\end{condition}

Condition (i) includes the required
non-informative censoring assumption and
the use of treatment information available
before each event time. Conditions (ii)
and (iii) ensure finite follow-up intensity
and a nonvanishing weighted risk set
\citep{andersen_statistical_1993,
fleming_counting_2005}. Each partial-likelihood contribution divides by a risk-set sum whose population counterpart is $c_t$; if $c_t$ approached zero, these denominators, and with them the score and information, would become unstable. 

\begin{proposition}[Consistency of the time-dependent Cox partial likelihood]
Under Assumptions~\ref{ass:sutva}--\ref{ass:history-invariance}
and Condition~\ref{cond:cox}, suppose
$\eta_0(x)=b(x)^\top\gamma_0$ is correctly specified
with a fixed-dimensional basis.
If the partial-likelihood information for
$(\gamma_0,\beta_0)$ is nonsingular and the standard
Cox moment conditions hold, then any regular maximizer
of Equation~\ref{eq:pl-tv-original-data} satisfies
\[
(\hat\gamma,\hat\beta)
\xrightarrow{p}(\gamma_0,\beta_0).
\]
\end{proposition}

The proposition is the standard consistency result for a Cox model with predictable time-dependent covariates. It follows from the martingale representation of the score and a uniform law of large numbers for the normalized partial likelihood \citep{andersen_statistical_1993, fleming_counting_2005}.  

The direct estimator is attractive when a low-dimensional model for $\eta_0$ is adequate. When baseline risk is learned by a flexible procedure, which typically converges more slowly than $n^{-1/2}$, its estimation error passes directly into $\hat\beta$, so $\hat\beta$ also converges too slowly for confidence intervals to attain nominal coverage. 
The next section changes the parametrization so that errors in the baseline-risk fit no longer pass directly into $\hat\beta$.

\section{Double Machine Learning Estimator}
\label{sec:DML}

\begin{figure}[t]
\centering
\includegraphics[width=\textwidth]
  {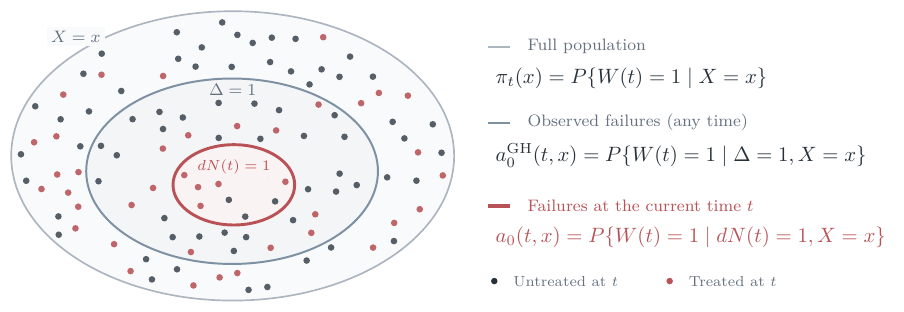}
\caption{
Nested populations underlying three propensity definitions at a fixed time \(t\) and covariate value \(X=x\). The ordinary propensity \(\pi_t(x)\) describes treatment status in the full population; \(a_0^{\mathrm{GH}}(t,x)\) restricts attention to observed failures across follow-up; and \(a_0(t,x)\) further conditions on failure at time \(t\). The latter is understood through an infinitesimal event-time interval and provides the required Neyman orthogonality for Cox partial-likelihood. Red and dark dots denote treated and untreated individuals, respectively. Population sizes and treatment proportions are schematic.
}
\label{fig:propensity-comparison}
\end{figure}

In this section, we develop TV-CSL, which allows the baseline-risk function to be estimated flexibly while protecting the treatment-effect estimator from first-order nuisance estimation error in the baseline-risk fit
\citep{chernozhukov2018double,gao2025estimating}. The key component is a risk-set propensity score, the probability of being treated among individuals who fail at time $t$, given their covariates. Intuitively, it is the proportion of treated individuals at each event time, among those who fail at that time. This is a nested population: smaller than the full population, and smaller than all failures over follow-up. Figure~\ref{fig:propensity-comparison} shows these nested populations.

We first show how centering by this propensity arises from the Cox partial likelihood (Section~\ref{sec:centering}). We then present the cross-fitted estimator and establish Neyman orthogonality and its convergence rate (Section~\ref{sec:neyman-ortho-theory}). Finally, we develop inference for the heterogeneous effect (Section~\ref{sec:inference}).

\subsection{Risk-set centering and event propensity}
\label{sec:centering}

We first rewrite the model so that treatment enters only through a centered indicator and all remaining dependence on $X$ is collected in one nuisance function.
Recall that
$\eta_1(x)-\eta_0(x)=p(x)^\top\beta_0$.
For any measurable function $a(t,x)$, define
\[
\nu_a(t,x)
=a(t,x)\eta_1(x)+\{1-a(t,x)\}\eta_0(x).
\]
Then
\[
\eta_w(x)
=\nu_a(t,x)
+\{w-a(t,x)\}p(x)^\top\beta_0,
\qquad w\in\{0,1\}.
\]

Although every choice of $a$ gives an exact
reparametrization, we seek one that makes
treatment-effect estimation insensitive to
first-order nuisance errors. As established
below, the required centering identity is
\begin{equation}
\label{eq:risk-set-orthogonality-identity}
E\left[
Y(t)e^{\eta_{W(t)}(X)}
\{W(t)-a_0(t,X)\}p(X)
\mid X
\right]=0.
\end{equation}
Under this identity, the centered regressor is uncorrelated, within each risk set and under hazard weighting, with every function of $X$, including the error in the fitted baseline risk \footnote{Reason why that error drops out at first order.}. 
To solve this equation, define
\[
R_w(t,x)
=P\{Y(t)=1,W(t)=w\mid X=x\},
\qquad w\in\{0,1\}.
\]
These are the treatment-specific risk-set
probabilities: the proportions of individuals
with covariates $x$ who remain at risk at $t$
in each treatment state.

Provided $p(x)\neq0$, the unique scalar centering solution on the event-risk support is
\begin{equation}
\label{eq:DINA-nuisance-time-varying}
\begin{split}
a_0(t,x)
&=
\frac{R_1(t,x)e^{\eta_1(x)}}
{R_1(t,x)e^{\eta_1(x)}
 +R_0(t,x)e^{\eta_0(x)}},\\
\nu_0(t,x)
&=\eta_0(x)+a_0(t,x)\tau_0(x).
\end{split}
\end{equation}

To the best of our knowledge, \(a_0(t,x)\) is a new type of propensity score in the literature. To aid interpretation, we now provide two ways to understand it.

\paragraph{A hazard-weighted change of measure.}

As previewed at the start of this section, the centering has a natural interpretation
as a change of measure. Using $c_t$
from Condition~\ref{cond:cox}, define
\begin{equation}
\label{eq:risk-set-measure}
dP_{R_t}(O)
=
\frac{Y(t)e^{\eta_{W(t)}(X)}}{c_t}
\,dP(O).
\end{equation}
Write $E_{R_t}$ for expectation under
$P_{R_t}$. Equation~
\eqref{eq:risk-set-orthogonality-identity}
is equivalent to
\[
E_{R_t}\left[
\{W(t)-a_0(t,X)\}p(X)\mid X
\right]=0,
\]
and therefore
\begin{equation}
\label{eq:event-propensity-measure}
a_0(t,x)
=E_{R_t}[W(t)\mid X=x].
\end{equation}

The measure $P_{R_t}$ first restricts
attention to subjects at risk at time $t$
and then weights them by their relative
event hazards. Consequently, $a_0(t,x)$
describes treatment among subjects
contributing to events at time $t$,
rather than treatment among all subjects
or merely among those still at risk.
This is the distribution underlying the
Cox partial-likelihood comparison.

\paragraph{Observable event propensity.}

By the counting-process intensity and
Bayes' rule, Equation~
\eqref{eq:event-propensity-measure}
also admits the characterization
\begin{equation}
\label{eq:event-propensity-main}
a_0(t,x)
=
P\{W(t)=1\mid dN(t)=1,X=x\}.
\end{equation}
The conditioning is understood as a
limiting probability over an infinitesimal
event-time interval. A brief proof: for subjects at risk at $t$ with treatment status $w$ and covariates $x$, the probability of failing in $[t,t+dt)$ is $e^{\eta_w(x)}\,d\Lambda_0(t)$, and the probability of being at risk with status $w$ is $R_w(t,x)$. Bayes' rule then gives
\[
P\{W(t)=1\mid dN(t)=1,X=x\}
=\frac{R_1(t,x)e^{\eta_1(x)}\,d\Lambda_0(t)}{R_1(t,x)e^{\eta_1(x)}\,d\Lambda_0(t)+R_0(t,x)e^{\eta_0(x)}\,d\Lambda_0(t)}=a_0(t,x),
\]
because $d\Lambda_0(t)$ cancels. A slightly more formal proof is in Appendix~\ref{sec:event-propensity}.

This representation is useful for
estimation: among observed failures,
we can regress treatment status
$W_i(U_i)$, a binary variable, on $(U_i,X_i)$, without
separately estimating $R_0$, $R_1$,
or the treatment-specific hazards. In our simulation and data analysis, we fit a logistic generalized additive model.

Although $a_0(t,x)$ differs from the
ordinary adoption propensity $\pi_t(x)$,
the overlap assumption also controls
event-time overlap.

\begin{lemma}[Event-time overlap]
Under Equation~\eqref{eq:ordinary-cox-hazard},
\[
\operatorname{logit}a_0(t,x)
=
\operatorname{logit}\pi_t(x)
+\tau_0(x)+\log\rho_t(x),
\]
where
\[
\rho_t(x)=
\frac{P(U\geq t\mid A<t,X=x)}
{P(U\geq t\mid A\geq t,X=x)}.
\]
Under Assumption~\ref{ass:overlap} and
$\|\tau_0\|_\infty<\infty$,
$a_0(t,x)$ is bounded away from zero
and one on $[t_0,\bar t]$.
\end{lemma}

Our centering differs from the modified
propensity of \cite{gao2025estimating}.
For baseline treatment $W$, their
construction uses
\begin{equation}
\label{eq:gao-hastie-propensity}
a^{\mathrm{GH}}(x)
=P(W=1\mid\Delta=1,X=x).
\end{equation}
Its direct staggered-adoption analogue is
\begin{equation}
\label{eq:gao-hastie-staggered}
a_0^{\mathrm{GH}}(t,x)
=P(A<t\mid\Delta=1,X=x).
\end{equation}
This pools observed failures
across all of follow-up, rather than conditioning on a failure at the particular time $t$. Because proportion of the treated in the risk set changes over follow-up, pooling observed failures across time gives the wrong centering in our problem, so errors in the baseline-risk fit can again pass into $\hat\beta$. In other words, $a_0^{\mathrm{GH}}$  need not satisfy Equation~\eqref{eq:risk-set-orthogonality-identity}
or yield a Neyman-orthogonal Cox
partial-likelihood score, as pointed out by \citet{gao2025estimating}.
This matters little when the treatment effect is small, since treatment then barely changes who fails when, but more when the effect is large (Appendix~\ref{sec:centering-isolation}). Our $a_0(t,x)$ avoids this problem at any effect size.

Figure~\ref{fig:propensity-comparison} illustrates the
nested populations underlying the three propensity definitions:
the full population with $X=x$, the subset experiencing
an observed failure ($\Delta=1$), and the smaller subset
experiencing a failure at time $t$ ($dN(t)=1$).

\subsection{Time-Varying Causal Survival Learner (TV-CSL)}
\label{sec:neyman-ortho-theory}

We now turn the centering into an estimator. TV-CSL replaces the treatment indicator in the Cox score by its centered version and estimates two nuisance functions by cross-fitting: the event propensity $a_0$ and the offset $\nu_0$, which collects baseline risk.

For a candidate centering function $a$,
let $q_a(t)=\{W(t)-a(t,X)\}p(X)$ be the centered regressor, with $q_0(t)=q_{a_0}(t)$, and let 
$\bar q(t;\beta,a,\nu)=
\frac{
E[Y(t)e^{\nu(t,X)+q_a(t)^\top\beta}q_a(t)]
}{
E[Y(t)e^{\nu(t,X)+q_a(t)^\top\beta}]
}$ be its average among subjects at risk at $t$.
The population partial-likelihood score is
\begin{equation}
\label{eq:population-score}
\Psi(\beta,a,\nu)
=
E\int_0^{\bar t}
\{q_a(t)-\bar q(t;\beta,a,\nu)\}
\,dN(t),
\end{equation}
which, as in the usual Cox score, compares each failing subject's regressor with the risk-set average. TV-CSL solves its cross-fitted sample analogue (Algorithm~\ref{alg:cox-tv}). Cross-fitting estimates the nuisances on one part of the sample and evaluates the score on the other, so a flexible learner's overfitting to a subject cannot bias that subject's own score contribution.

\begin{algorithm}[tp]
\caption{Risk-set-orthogonalized TV-CSL}
\label{alg:cox-tv}
\begin{algorithmic}[1]
\State \textbf{Input:} $\{O_i\}_{i=1}^n$, basis $p(\cdot)$, subject-level folds $I_1,\ldots,I_K$ with $n_k=|I_k|$, and horizon $\bar t$.
\For{$k=1,\ldots,K$}
\State \textit{Propensity.} On $I_k^c$, regress treatment status at failure, $W_i(U_i)$, on $(U_i,X_i)$ among observed failures to obtain $\hat a^{(-k)}$ (Equation~\eqref{eq:event-propensity-main}).
\State \textit{Offset.} On $I_k^c$, fit a preliminary time-dependent survival model for $\eta_0,\eta_1$ (including any preliminary $\tilde\beta$) and form $\hat\nu^{(-k)}$ by Equation~\eqref{eq:DINA-nuisance-time-varying}.
\State \textit{Centering.} For $i\in I_k$, subtract the fitted propensity from treatment status,
\begin{equation}
\label{eq:fold-design}
\hat q_i(t)=\{W_i(t)-\hat a^{(-k)}(t,X_i)\}p(X_i),
\end{equation}
and set the linear predictor $\hat\theta_i(t;\beta)=\hat\nu^{(-k)}(t,X_i)+\hat q_i(t)^\top\beta$.
\State \textit{Risk-set averages.} Within $I_k$, compute
\begin{equation}
\label{eq:fold-risk-sums}
\hat S_k^{(m)}(t;\beta)=\frac1{n_k}\sum_{j\in I_k}Y_j(t)e^{\hat\theta_j(t;\beta)}\hat q_j(t)^{\otimes m},
\quad m=0,1,2,
\end{equation}
with $q^{\otimes0}=1$, $q^{\otimes1}=q$, $q^{\otimes2}=qq^\top$, and the hazard-weighted average $\hat{\bar q}_k=\hat S_k^{(1)}/\hat S_k^{(0)}$.
\EndFor
\State \textit{Score.} Holding $\hat\nu$ fixed as an offset, solve for $\hat\beta$
\begin{equation}
\label{eq:fold-score}
0=\frac1{n}\sum_{k=1}^K\sum_{i\in I_k}
\int_0^{\bar t}
\{\hat q_i(t)-\hat{\bar q}_k(t;\hat\beta)\}
\,dN_i(t).
\end{equation}
\State \textbf{Output:} $\hat\beta$ and $\hat\tau(x)=p(x)^\top\hat\beta$.
\end{algorithmic}
\end{algorithm}

We now establish theoretical properties of the estimator. Denote the treatment information matrix as
\begin{equation}
\label{eq:information-J}
J=
\int_0^{\bar t}
c_tE_{R_t}[q_0(t)q_0(t)^\top]
\,d\Lambda_0(t).
\end{equation}

We impose Condition~\ref{cond:cox}
and the following additional conditions.
The event-weighted norm
$\|\cdot\|_{2,\mu}$ is defined in
Appendix~\ref{sec:dml-technical}.

\begin{condition}[DML regularity]
\label{cond:dml}
\begin{enumerate}[(i)]

\item $E\|p(X)\|^4<\infty$, and the
eigenvalues of $J$ are bounded away
from zero and infinity.

\item The cross-fitted nuisance estimates
satisfy
\begin{equation}
\label{eq:nuisance-rates}
\|\hat a-a_0\|_{2,\mu}
=O_p(r_{a,n}),
\qquad
\|\hat\nu-\nu_0\|_{2,\mu}
=O_p(r_{\nu,n}),
\end{equation}
where $r_{a,n}+r_{\nu,n}=o(1)$.

\end{enumerate}
\end{condition}

Additional technical conditions are stated
in Appendix~\ref{sec:dml-technical}.
Cross-fitting permits flexible nuisance
learners without requiring Donsker
conditions on their function classes.

Our first result shows what the centering buys: with $a=a_0$, small errors in either nuisance change the score only at second order. 
\begin{proposition}[Neyman orthogonality]
\label{prop:orthogonality}
Suppose Conditions~\ref{cond:cox}
and~\ref{cond:dml-technical}(ii) hold.
At the true parameters,
\[
\Psi(\beta_0,a_0,\nu)=0
\]
for every admissible $\nu$. Moreover,
for all admissible directions $h_a,h_\nu$,
\[
D_a\Psi(\beta_0,a_0,\nu_0)[h_a]=0,
\qquad
D_\nu\Psi(\beta_0,a_0,\nu_0)[h_\nu]=0.
\]
\end{proposition}

The proof is given in
Appendix~\ref{sec:proof-prop:orthogonality}.
The result is stronger than ordinary
first-order insensitivity to $\nu$:
when $a=a_0$, the population score
vanishes for every admissible $\nu$.
This removes the pure $r_{\nu,n}^2$
term from the nuisance remainder. In words, errors in a flexibly learned baseline risk no longer pass directly into $\hat\beta$.

The next result translates orthogonality into a rate: the error of $\hat\beta$ is the usual $n^{-1/2}$ term plus a remainder driven by products of nuisance errors.
\begin{proposition}[DML convergence rate]
\label{prop:dml-rate}
Suppose Conditions~\ref{cond:cox},
\ref{cond:dml}, and
\ref{cond:dml-technical} hold.
Then
\[
\|\hat\beta-\beta_0\|
=
O_p\left\{
n^{-1/2}
+r_{a,n}(r_{a,n}+r_{\nu,n})
\right\}.
\]
In particular, root-$n$ consistency
follows if
\[
\sqrt n\,r_{a,n}
(r_{a,n}+r_{\nu,n})\to0.
\]
A sufficient condition is
$r_{a,n}=o(n^{-1/4})$ and
$r_{\nu,n}=o(n^{-1/4})$.
\end{proposition}

The proof is in
Appendix~\ref{sec:proof-dml-rate}.
The rate is asymmetric: accurate
event-time centering allows slower
convergence of the outcome nuisance,
but not conversely. Because the remainder is a product, each nuisance may converge more slowly than $n^{-1/2}$; it suffices that both converge faster than $n^{-1/4}$, which many flexible learners achieve under smoothness or sparsity conditions. By contrast, the direct Cox estimator has error of order $r_{\nu,n}$ itself.

\subsection{Inference with TV-CSL}
\label{sec:inference}

We now turn the convergence rate into confidence intervals. We first show that $\hat\beta$ is asymptotically normal and then estimate its variance.

Let $M_i(t) = N_i(t) -\int_0^t Y_i(s)e^{\eta_{W_i(s)}(X_i)} \,d\Lambda_0(s)$ be the event martingale of subject $i$, the observed event count minus its model-predicted risk, and let $q_i^0(t)$ be the
subject-specific realization of $q_0(t)$. Define the influence contribution and its variance,
\[
\psi_i=\int_0^{\bar t}
q_i^0(t)\,dM_i(t),
\qquad
\Sigma=E(\psi_i\psi_i^\top),
\qquad
\Omega=J^{-1}\Sigma J^{-1}, 
\]
with $J$ from Equation~\eqref{eq:information-J}.
Unlike the population score in Equation~\eqref{eq:population-score}, $\psi_i$ integrates the centered regressor against the event martingale rather
than the observed counting process, so each contribution has mean zero.

Proposition~\ref{prop:dml-rate} gives a rate but not a distribution. Under a slightly stronger rate condition, the next result shows that $\hat\beta$ behaves like an average of the $\psi_i$.
\begin{proposition}[Asymptotic normality]
\label{prop:dml-clt}
Suppose the conditions of
Proposition~\ref{prop:dml-rate} hold and
\[
\sqrt n\,r_{a,n}
(r_{a,n}+r_{\nu,n})\to0.
\]
Then
\[
\sqrt n(\hat\beta-\beta_0)
=
J^{-1}\frac1{\sqrt n}
\sum_{i=1}^n\psi_i+o_p(1)
\ \rightsquigarrow\ N(0,\Omega).
\]
\end{proposition}

The proof is given in
Appendix~\ref{sec:proof-dml-clt}.
This is the Cox partial-likelihood
analogue of best-linear-predictor
inference in
\cite{semenova2021debiased}. The asymptotic variance $\Omega$ has a sandwich form: $J$ measures how sharply the score identifies $\beta_0$, and $\Sigma$ measures the variability of the individual contributions.

We estimate $J$ and $\Sigma$ by fold-level plug-ins that reuse the risk-set sums from Algorithm~\ref{alg:cox-tv}. The full procedure is shown in Algorithm~\ref{alg:variance}. The resulting sandwich standard error for $\hat\beta_j$ is $\{\hat\Omega_{jj}/n\}^{1/2}$. When the Cox model is correctly specified and the nuisance conditions of Proposition~\ref{prop:dml-clt} hold, $\Sigma=J$, which gives the simpler information-based standard error $\{(\hat J^{-1})_{jj}/n\}^{1/2}$. We report both in the simulations. For any covariate value $x$, the $(1-\alpha)$ confidence interval for the heterogeneous log hazard ratio is

\[
p(x)^\top\hat\beta
\ \pm\
z_{1-\alpha/2}
\left\{
\frac{p(x)^\top\hat\Omega p(x)}n
\right\}^{1/2}.
\]

\begin{algorithm}[t]
\caption{Variance estimation for TV-CSL}
\label{alg:variance}
\begin{algorithmic}[1]
\State \textbf{Input:} $\hat\beta$ and, for each fold $k$, $\hat q_i$, $\hat\theta_i$, $\hat S_k^{(m)}$, and $\hat{\bar q}_k$ from Algorithm~\ref{alg:cox-tv}.
\For{$k=1,\ldots,K$}
\State \textit{Information.} Compute the hazard-weighted risk-set covariance
\[
\hat V_k(t)=
\frac{\hat S_k^{(2)}(t;\hat\beta)}
{\hat S_k^{(0)}(t;\hat\beta)}
-\hat{\bar q}_k(t;\hat\beta)
\hat{\bar q}_k(t;\hat\beta)^\top.
\]
\State \textit{Residuals.} For $i\in I_k$, compute the Breslow increment and martingale residual
\[
d\hat\Lambda_{0,k}(t)
=
\frac{\sum_{l\in I_k}dN_l(t)}
{\sum_{j\in I_k}Y_j(t)e^{\hat\theta_j(t;\hat\beta)}},
\qquad
d\hat M_i(t)
=
dN_i(t)
-Y_i(t)e^{\hat\theta_i(t;\hat\beta)}
\,d\hat\Lambda_{0,k}(t),
\]
and the influence contribution $\hat\psi_i=\int_0^{\bar t}\{\hat q_i(t)-\hat{\bar q}_k(t;\hat\beta)\}\,d\hat M_i(t)$.
\EndFor
\State \textit{Combine.} Set
\[
\hat J=
\frac1n\sum_{k=1}^K
\sum_{i\in I_k}
\Delta_i\mathbf 1(U_i\leq\bar t)\hat V_k(U_i),
\qquad
\hat\Sigma
=\frac1n\sum_{i=1}^n
\hat\psi_i\hat\psi_i^\top,
\qquad
\hat\Omega
=\hat J^{-1}\hat\Sigma\hat J^{-1}.
\]
\State \textbf{Output:} $\hat\Omega$, standard errors, and confidence intervals for $\tau_0(x)$.
\end{algorithmic}
\end{algorithm}

\section{Simulation Study}

In the simulation study, we investigate when TV-CSL improves on simpler time-dependent Cox estimators and when it does not. We have two conclusions. First, for point estimation, TV-CSL achieves lower mean squared error when the baseline risk function $\eta_0$ is complex and the sample size is sufficiently large, because a flexible fit of $\eta_0$ converges slowly and the direct Cox estimator inherits its error at first order, whereas TV-CSL is affected only through products of nuisance errors. When baseline risk is simple, or when the sample size is small, direct Cox estimation is good enough, because there is no first-order bias to remove and TV-CSL pays the variance cost of estimating extra nuisance functions.
Second, for inference, TV-CSL orthogonalization is necessary to achieve reliable coverage when $\eta_0$ is complex, because the bias of the direct estimator does not shrink as fast as its standard error; when a low-dimensional Cox model is correctly specified, direct Cox intervals are already valid. The rest of the section provides details.

\subsection{Simulation Design}
\label{sec:simulation-design}
The simulation design builds in setup discussed in theory, that is, adoption timing that is confounded with outcome risk, and baseline risk that may be too complex for a low-dimensional Cox model. For $i=1,\ldots,n$, let $X_i=(X_{i1},X_{i2},X_{i3})^\top$ draw from independent $\operatorname{Normal}(0,1)$. Adoption time depends on $x_2$ and $x_3$,
\[
A_i\mid X_i=x\sim\operatorname{Exp}\{r(x)\},
\qquad r(x)=e^{x_2+x_3}.
\]
Given adoption at $a$, the event hazard is
\[
h(t\mid a,x)=t\exp\{\eta_0(x)+\mathbf 1(a<t)\tau_0(x)\},
\qquad \tau_0(x)=p(x)^\top\beta_0=x_1+x_2+x_3,
\]
where $p(x)=(1,x_1,x_2,x_3)^\top$ and $\beta_0=(0,1,1,1)^\top$. The treatment effect is heterogeneous and lies in the span of $p(x)$. This also means $x_2, x_3$ are confounders. 

We consider two baseline log-hazard functions, varying complexity of outcome risk. The first is linear,
\[
\eta_0^L(x)=2.5(x_1+x_2/2+x_3/3),
\]
so a low-dimensional Cox model fits it exactly.
The second is non-linear and non-additive:
\[
\eta_0^{NL}(x)=-\tfrac34\varsigma(x_1)\varsigma(x_2),
\qquad
\varsigma(z)=\frac{2}{1+e^{-12(z-1/2)}},
\]
so it must be learned flexibly. Comparing the two scenarios therefore compares a setting in which orthogonalization is not needed with one in which it is.

Independent censoring is generated as $C_i=\min(20,\widetilde C_i)$ with $\widetilde C_i\sim\operatorname{Exp}(0.1)$. We use $n\in\{200,500,1000,2000\}$ and $1000$ Monte Carlo replications per setting. The same sample sizes, treatment process, treatment effect, and censoring mechanism are used in both baseline-risk scenarios.

\subsection{Methods for Comparison}
\label{sec:methods-comparison}

We compare TV-CSL with two single-stage Cox estimators, \emph{S-Lasso} and \emph{S-Cox}, where ``single-stage'' means that no propensity model is used. \emph{S-Lasso} fits the outcome model with a cross-validated lasso penalty, whereas \emph{S-Cox} fits the same model structure using an unpenalized Cox model.

All methods use two specifications for the baseline-risk function $\eta_0$: a linear specification and a complex specification containing natural splines, squared terms, and pairwise interactions. These are fitted model classes, distinct from the two DGPs described above. Thus, we consider two DGPs crossed with two outcome-model specifications. All methods use a linear model for $\tau_0$. TV-CSL uses \emph{S-Lasso} as the outcome nuisance model.

Furthermore, TV-CSL uses one of three propensity specifications:
\begin{itemize}
\item \emph{Intercept-only} uses a constant propensity and is deliberately misspecified.
\item \emph{Marginal propensity} is the direct staggered-adoption extension of \cite{gao2025estimating} in Equation~\ref{eq:gao-hastie-staggered}. Among subjects with $\Delta_i=1$, it fits a Cox model for treatment initiation with response
$\operatorname{Surv}\{\min(A_i,U_i),\mathbf 1(A_i<U_i)\}$ and predictors $X_i$.
\item \emph{Time-varying propensity} estimates our $a_0(t,x)$ directly by fitting the event regression
$\operatorname{logit}a_0(t,x)=s(t)+x_1+x_2+x_3$ on
$\{(U_i,X_i,W_i(U_i)): \Delta_i=1\}$.
\end{itemize}

Thus, TV-CSL has six fitted variants per DGP: two outcome specifications crossed with three propensity specifications. Each comparison tests one question: S-Cox versus TV-CSL asks whether orthogonalization pays off relative to direct estimation; the intercept-only propensity asks what happens when the centering is wrong; and marginal versus time-varying propensity asks whether conditioning on failure at time $t$, rather than at any time, matters in practice.

We do not include causal survival forests \citep{cui2023estimating} or other survival learners for heterogeneous effects. These methods require treatment to be fixed at baseline and target effects on restricted mean survival time or survival probabilities. Applying them here would require either classifying units by eventual adoption, which Section~6.2 shows to be biased, or using a landmark design with a different estimand. Such a comparison would therefore confound the choice of estimand with the handling of treatment timing.

We report mean squared error (MSE), $\|\hat\beta-\beta_0\|^2$, and empirical 95\% coefficient-wise coverage, averaged over the three treatment-effect coordinates. For TV-CSL, coverage is calculated using both the information-based and sandwich standard errors from Section~\ref{sec:DML}. The single-stage estimators use the usual Cox information standard errors.

\subsection{Results}

\begin{figure}[tp]
\centering
\includegraphics[width=\textwidth,height=0.78\textheight,keepaspectratio]{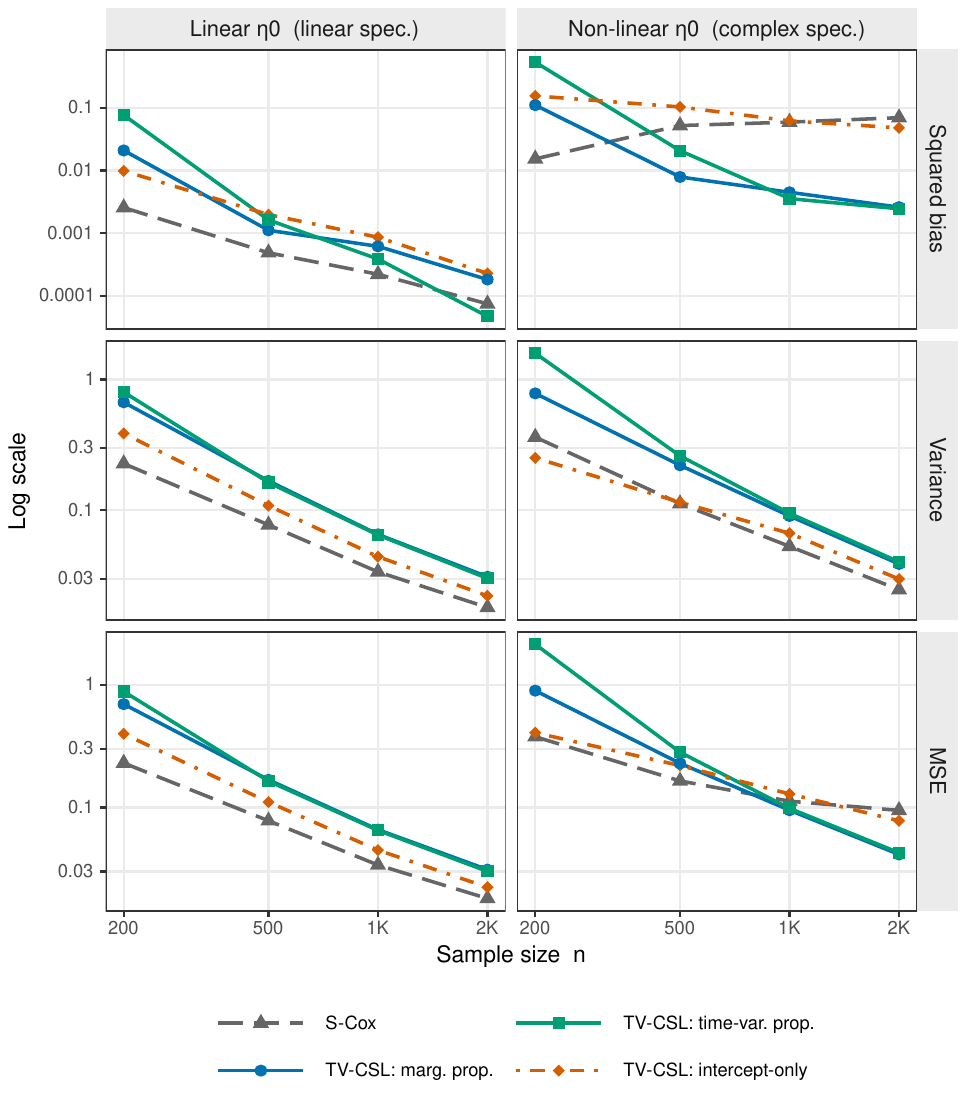}
\caption{Squared bias (top), variance (middle), and MSE (bottom) of $\hat\beta$ by sample size, for linear (left) and non-linear (right) $\eta_0$.}
\label{fig:bias-variance-mse}
\end{figure}

\begin{figure}[ht]
\centering
\includegraphics[width=\textwidth]{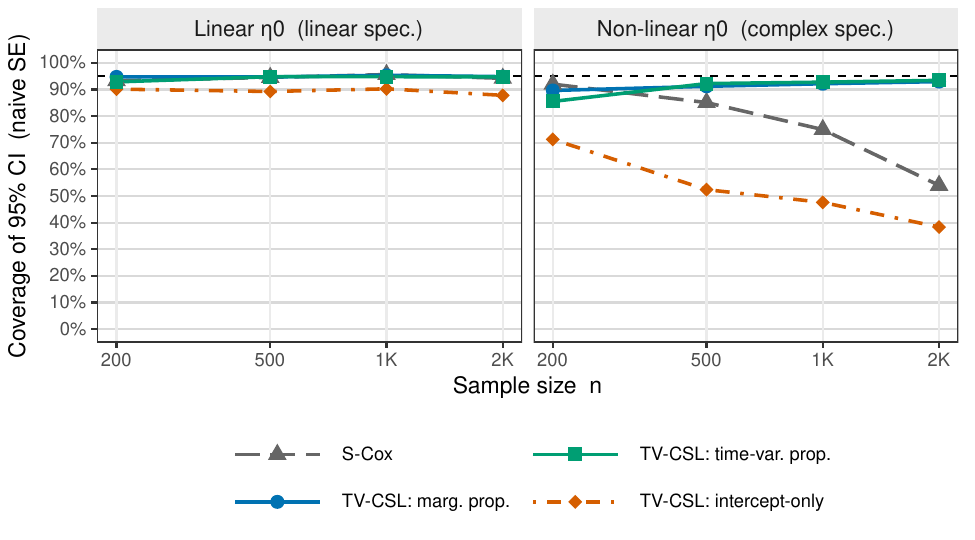}
\caption{Empirical 95\% coverage using information-based standard errors, averaged over the three treatment-effect coefficients. The dashed line marks 0.95. Under non-linear $\eta_0$, near-nominal coverage is obtained only by TV-CSL with the marginal or direct time-varying propensity.}
\label{fig:coverage}
\end{figure}

The main results are presented in Figures~\ref{fig:bias-variance-mse} and~\ref{fig:coverage}. In these figures, we omit S-Lasso because S-Cox performs as well or better in the settings shown. Appendix~\ref{sec:simulation-tables} reports MSE and coverage for every fitted outcome specification.

We first discuss MSE and its decomposition into squared bias and variance (Figure~\ref{fig:bias-variance-mse}).
When $\eta_0$ is linear, direct Cox estimation performs better than the TV-CSL methods across all sample sizes. This is expected: when all models are correctly specified, the estimators are nearly unbiased, so the comparison is decided by variance and the simpler model, S-Cox, performs better in finite samples. When $\eta_0$ is non-linear, the picture changes. The complex specification cannot fully recover $\eta_0$, and S-Cox inherits this error: its squared bias does not shrink with $n$, so its MSE levels off. TV-CSL with the marginal or time-varying propensity is protected from this error, so its bias and variance both fall with $n$. TV-CSL has larger variance at small samples, so S-Cox has lower MSE below about $n=1000$; at $n=2000$, TV-CSL's MSE is less than half that of S-Cox. The intercept-only propensity does not center treatment correctly, so it keeps substantial bias.

We next discuss coverage as shown in Figure~\ref{fig:coverage}. Under linear $\eta_0$, all procedures achieve coverage close to the nominal level except TV-CSL with the intercept-only propensity. Under non-linear $\eta_0$, only TV-CSL with the marginal or time-varying propensity achieves coverage close to the nominal level. The poor coverage of Cox reflects the persistent bias of the misspecified baseline-risk fit, whereas the poor coverage of TV-CSL with the intercept-only propensity results from misspecification of the propensity model, i.e., the conditions of Proposition~\ref{prop:dml-clt} are not satisfied. For both Cox and TV-CSL with the intercept-only propensity, coverage deteriorates as the sample size increases because the estimators converge more precisely to a biased target. We did not run this comparison beyond $n=2000$. Appendix~\ref{sec:centering-isolation} reports TV-CSL alone at $n=4000$; there, under the non-linear $\eta_0$, its remaining bias is driven by the outcome-model error.

In Figure~\ref{fig:coverage}, coverage is based on the information-based variance estimator. We find that information-based TV-CSL intervals are closer to nominal coverage than the empirical sandwich intervals at smaller sample sizes, while the two become closer as $n$ increases.

Among the TV-CSL specifications, marginal and time-varying centering give similar MSE and coverage, whereas the intercept-only specification is clearly worse because of propensity-score misspecification. In theory, time-varying centering is the centering implied by the orthogonality condition, whereas marginal centering is only an approximation. Their similar performance in the main simulation appears to arise because both estimated propensities differ from $a_0$: the marginal propensity by definition, and the time-varying propensity because its estimate relies on a working model. Under the non-linear $\eta_0$, both errors are multiplied by a large outcome-model error. In Appendix~\ref{sec:centering-isolation}, we therefore consider a simulation designed to isolate the centering difference. There, exact time-varying centering has smaller bias and better coverage than marginal centering, and the difference increases with the magnitude of the treatment effect.


\section{Data Analysis: Copilot Adoption}
\label{sec:copilot-application}

Generative-AI tools are increasingly available in professional services, but trying a tool once does not imply that it becomes part of regular work. We use lawyer-level data from a global law firm to estimate whether, and by how much, first observed use of the firm's internally developed AI assistant (the treatment), changes the rate of becoming a regular user of a second AI tool, Microsoft Copilot (the outcome). We also examine whether this effect differs between junior and senior lawyers. This question is important because firms increasingly deploy several complementary AI tools, and whether these tools are complements or substitutes is an open question: if experience with one tool makes another easier to adopt, investment in one raises the return to the other, whereas tools serving overlapping needs may crowd each other out. Evidence on which pattern holds can inform investments in product integration, training, and employee adoption.

Time is measured in weeks from 6 October 2024, when Copilot telemetry begins. We define regular Copilot use as the first week with at least five recorded Copilot actions, or prompts, across applications such as Outlook and PowerPoint. The threshold was chosen through an exploratory examination of weekly usage patterns. It is an operational measure of engagement: reaching it once does not imply sustained use, and the available action counts do not distinguish the quality or usefulness of those actions. Lawyers who do not reach the threshold during observable follow-up are censored.

The analysis has two goals. First, we show how treating eventual use of the AI assistant as a baseline treatment, rather than accounting for its adoption date, changes the estimates. Second, using first use of the AI assistant  as a time-varying treatment, we compare the time-varying Cox model with TV-CSL. Throughout this section, the main effect measure we report is the hazard ratio. A hazard ratio of \(h\) means that lawyers who had started using the AI assistant became regular Copilot users at approximately \(h\) times the instantaneous rate, at any given time, of comparable lawyers who had not yet started using the AI assistant.

\subsection{Study Population, Variables for the Analysis and Summary Statistics}

The unit of analysis is a lawyer; first use of the AI assistant is the treatment, and regular Copilot use is the event. Lawyers with no observed use of the AI assistant after time zero remain untreated during follow-up. When initiation and regular Copilot use occur in the same week, the primary convention records the event as untreated. Follow-up ends at week 81, with earlier censoring at the last week the lawyer appears in the Copilot telemetry, including weeks with no Copilot use. We assume this censoring is non-informative given the adjustment covariates; telemetry presence approximates employment but can end earlier for lawyers who become inactive in the system.~\footnote{These conventions follow the framework of Section~\ref{sec:causal-target}: recording a same-week adoption and event as untreated keeps treatment status determined before the event, matching $W_i(t)=\mathbf 1(A_i<t)$, and the censoring assumption is the non-informative censoring in Condition~\ref{cond:cox}(i).}

The adjustment covariates are rank, office, practice group, weekly working hours, and prior use of the AI assistant. Rank records a lawyer's professional level, office records the lawyer's location, and practice group records the lawyer's area of legal practice. Weekly working hours are averaged over each lawyer's first four observed weeks. Seniority, derived from rank, is the treatment-effect modifier and defines the effect basis $p(x)$: Partners, Counsel, Senior Associates, and Mid-Level Associates are classified as senior, whereas Junior Associates and Trainees are classified as junior.

Of the whole 3,874 lawyers in the dataset, we exclude 551 who were already regular Copilot users at the start of follow-up, because their transition to regular use is not observed. This leaves 3,323 lawyers. Table~\ref{tab:summary-stat-copilot} reports the summary statistics for the 3,323 lawyers. We report only a subset of the summary statistics to prevent the company from being identified.

For the primary analysis, we further exclude 1,296 lawyers who used the AI assistant between 23 June 2024, when the AI assistant telemetry begins, and time zero, so that the analysis includes only new users of the AI assistant. We also exclude one lawyer who became a regular user in their entry week and so contributes no time at risk. The primary population is 2,026 lawyers, of whom 1,891 became regular Copilot users during follow-up. As a sensitivity analysis, we analyse all 3,322 eligible lawyers (the 3,323 minus the same one lawyer). There, we use prior use of the AI assistant as a binary adjustment variable in fitting nuisance models.


Among the 1,891 regular-use events in this subgroup, 729 (38.6\%) occurred after the first use of AI assistant and 1,162 (61.4\%) occurred before or in the same week as first use. Both treatment states were represented within each event-time and seniority group. This means the treatment variable in the dataset is proper for time-varying effect estimation. 

\begin{table}[tp]
\centering
\footnotesize
\setlength{\tabcolsep}{4pt}
\renewcommand{\arraystretch}{0.92}
\caption{Summary Statistics for the Copilot Adoption Analysis. To reduce the risk of identifying the company, rank and office are reported in coarsened categories; offices are grouped into the eight largest offices and all remaining offices.}
\label{tab:summary-stat-copilot}
\resizebox{\linewidth}{!}{%
\begin{tabular}{llr}
\toprule
Characteristic & Category or summary & Overall \\
\midrule
Seniority & Senior, count (\%) & 1,809 (54.4) \\
          & Junior, count (\%) & 1,514 (45.6) \\
\midrule
Rank (\%) & Partner & 20.0 \\
          & Associate & 55.0 \\
          & Other & 25.0 \\
\midrule
Office (\%) & Eight largest offices & 67.9 \\
            & Other offices & 32.1 \\
\midrule
Working hours, first four observed weeks
    & Mean (SD); median [IQR]
    & 39.0 (12.8); 40 [32, 46] \\
\bottomrule
\end{tabular}%
}
\end{table}

\subsection{Effect of Ignoring Treatment Time}

In this section, we compare two Cox models using the same cohort. The first treats eventual observed use of the AI assistant as a fixed treatment variable. A subject is classified as treated if they receive treatment at any point, regardless of when treatment occurs. A control is a subject who receives no treatment before the event of becoming a regular user. The second uses the time-varying treatment process in Equation~\ref{eq:ordinary-cox-hazard}, with first use of the AI assistant as adoption.

Table~\ref{tab:copilot-ignore-time} shows that the two models give very different answers.
Both models find a significant effect for senior and junior lawyers, but the magnitudes are different: compared to the time-varying model, in the fixed treatment model, seniors have a 1.29 times effect in terms of hazard ratio (point estimate: 2.11 divided by 1.63),
while juniors have a 1.69 times effect (point estimate: 2.14 divided by 1.27).
Most importantly, the difference between junior and senior is significant in the time-varying model (ratio 0.78, $p = 0.015$) but not in the fixed treatment model (ratio 1.01, $p = 0.934$).

\input{tables/tab_copilot_ignore_time}

\subsection{TV-CSL Results and Sensitivity Analyses}

In this section, we apply three estimators to the data: the time-varying Cox model from the previous section and two TV-CSL specifications, one using the Gao--Hastie propensity in Equation~\ref{eq:gao-hastie-staggered} and the other using the time-varying propensity in Equation~\ref{eq:event-propensity-main}. Nuisance models are cross-fitted at the lawyer level.

The results are reported in Table~\ref{tab:copilot-tvcsl}. The point estimates are similar across the two TV-CSL specifications, as expected, although their confidence intervals differ. Qualitatively, both specifications suggest a positive association between the AI assistant use and becoming a regular Copilot user, with a larger estimated effect among senior lawyers. Quantitatively, we focus on TV-CSL with the time-varying propensity because the simulation study indicates that this specification performs well. At any given time, senior lawyers who had started using the AI assistant  became regular Copilot users at about 1.5 times the rate of comparable non-users (95\% CI: 1.3--1.8), while junior lawyers did so at about 1.2 times the rate (95\% CI: 1.0--1.5).

We also conducted sensitivity analyses, reported in Table~\ref{tab:copilot-sensitivity} in the Appendix. The qualitative conclusion is unchanged when we vary how same-week adoption and event times are handled, although the estimated effects become larger. If lawyers who first used the AI assistant in the same week that they became regular Copilot users are counted as having used the AI assistant first, the estimated hazard ratios increase to approximately 1.9 for senior lawyers and 1.8 for junior lawyers. The two analysis populations agree for senior lawyers but not for junior lawyers. In the all-lawyers cohort, under the primary tie convention, the senior hazard ratio is about 1.55, close to the primary estimate, whereas the junior hazard ratio is about 1.05 with a confidence interval that includes one. The positive effect for junior lawyers is therefore not present once prior users of the AI assistant are included.

\input{tables/tab_copilot_tvcsl}

If the measured adjustment variables are sufficient to control confounding and the identifying assumptions in Section~\ref{sec:causal-target} hold, the time-varying TV-CSL estimates have a causal interpretation: first observed use of the AI assistant increases the instantaneous rate of becoming a regular Copilot user for both senior and junior lawyers. One possible explanation is that the AI assistant provides lawyers with an accessible introduction to AI-assisted work, making subsequent regular Copilot use easier. In the terms of our motivation, the two tools behave as complements. The point estimate is larger for senior lawyers, although under TV-CSL with the time-varying propensity the confidence interval for the junior/senior ratio includes one (Table~\ref{tab:copilot-tvcsl}). Shared enthusiasm for new technology that our covariates do not capture would produce the same pattern, so we regard the mechanism as a hypothesis.

\section{Conclusion}
We have connected a staggered-adoption causal design to a survival model for a once-observed event time. Under no anticipation, invariance to adoption history, unconfounded adoption time, and overlap, the time-dependent Cox contrast has a causal interpretation as a heterogeneous log hazard ratio. For estimation, the key object is not the ordinary adoption propensity but the probability of prior treatment among subjects who fail at the current event time. Centering by this risk-set modified propensity yields a Neyman-orthogonal partial-likelihood score, a second-order nuisance remainder, and influence-function-based pointwise inference. More broadly, the hazard-weighted change of measure explains why treatment centering in survival analysis must account for both the contemporaneous risk set and the relative event hazards. This perspective may open the door to developing orthogonal estimating equations for other causal survival problems.

The simulations give a practical qualification to the theory. When baseline risk is simple, direct Cox estimation is more precise and there is little reason to estimate another nuisance. When baseline risk is complex, the cost of orthogonalization is eventually outweighed by its protection against first-stage error. In that setting, valid event-time centering is also important for inference: a deliberately misspecified constant propensity has both higher MSE and severe undercoverage.

The framework applies whenever treatment is adopted once and both treated and not-yet-treated subjects can experience the event at the same analysis times. The Copilot study shows why this design condition must be checked before estimation, while the Stanford Heart Transplant data in Appendix~\ref{sec:stanford-appendix} provide the classical setting in which events may occur both before and after adoption.

\subsection{Limitations and Future Work}
\label{sec:limitations}

The causal interpretation depends on stronger conditions than the statistical Cox fit alone. SUTVA may fail when an intervention has materially different versions, such as transplant procedures that vary across surgeons or hospitals. Unconfoundedness requires measurement of the variables that jointly determine adoption and the event. In transplantation, these include disease severity, comorbidities, organ compatibility, geography, and allocation rules \citep{cascino_comparison_2022, khush_donor_2013, parker_association_2019, power_contemporary_2024}. Overlap can also be poor when such rules nearly determine who is treated at a given time. These are design limitations and cannot be repaired by orthogonalization.

No anticipation and invariance to adoption history are also substantive restrictions. The latter excludes duration-dependent effects. For example, mortality risk may increase immediately after transplantation because of surgical complications or acute rejection and then decline among patients who survive the early period \citep{lipkova2022deep}. Such a pattern requires treatment-time or duration-specific effects rather than the single post-adoption hazard ratio used here. The present framework also does not address time-dependent confounders affected by prior treatment. More generally, if the treatment effect varies with calendar time or time since adoption, the proportional-hazards restriction is misspecified and the resulting HLHR should be interpreted as a working summary; extending TV-CSL to a time-varying contrast \(\tau(t,x)\) is left for future work.

Our finite-dimensional, prespecified basis $p(x)$ is another limitation. It permits interpretable transformations and interactions but does not adapt non-parametrically to an unknown effect surface. Future work could develop an information-weighted projection target, uniform confidence bands, and tests of which subgroups benefit, extending related work for continuous outcomes \citep{dukes2024nonparametric}. It would also be useful to study sensitivity analysis for unmeasured adoption-time confounding and diagnostics for event-risk overlap. 

Our estimator of the event propensity $a_0(t,x)$ is also imperfect. The logistic working model used here is a practical approximation, and the simulations suggest that its error can limit the benefit of time-varying centering when the outcome model is poor. Developing flexible estimators of $a_0$ that respect its structure as a treated fraction among subjects failing at $t$ is an important direction for future work.

\bibliographystyle{plainnat}
\bibliography{refs}

\appendix

\section{Direct estimation of the event propensity}
\label{sec:event-propensity}

This section derives three equivalent interpretations
of the event propensity $a_0(t,x)$: the treatment
probability among individuals experiencing an event
at time $t$, the treatment probability under the
hazard-weighted risk-set measure $P_{R_t}$, and
an adjustment of the ordinary adoption propensity
for survival selection and relative event hazards.
We then describe its direct estimation and derive
the analytical oracle used in Appendix~\ref{sec:centering-isolation}.

\subsection{Event propensity and change of measure}

We establish the observable characterization
in Equation~\eqref{eq:event-propensity-main}
and its equivalence to the
hazard-weighted measure $P_{R_t}$.

For $h>0$, write
$\Delta_hN(t)=N(t+h)-N(t)$.
Under the counting-process intensity,
at regular event times,
\begin{align}
&P\{W(t)=w,\Delta_hN(t)=1\mid X=x\}
\nonumber\\
&\quad=
R_w(t,x)\lambda_0(t)e^{\eta_w(x)}h
+o(h).
\label{eq:appendix-event-joint}
\end{align}
Here $o(h)/h\to0$ as $h\downarrow0$;
the expansion assumes local continuity
and excludes simultaneous adoption and
failure jumps.

Applying Bayes' rule and cancelling
the common factor $\lambda_0(t)h$
gives, whenever $\lambda_0(t)>0$,
\begin{equation}
\label{eq:event-propensity-characterization}
\begin{split}
a_0(t,x)
&=\lim_{h\downarrow0}
P\{W(t)=1\mid
\Delta_hN(t)=1,X=x\}\\
&=\frac{R_1(t,x)e^{\eta_1(x)}}
{\sum_{w=0}^1R_w(t,x)e^{\eta_w(x)}}.
\end{split}
\end{equation}
This proves
Equation~\eqref{eq:event-propensity-main}.

The equivalence extends to the joint
distribution of pre-event characteristics.
For any measurable covariate set $B$,
\[
\begin{aligned}
&\lim_{h\downarrow0}
P\{W(t)=w,X\in B\mid
\Delta_hN(t)=1\}\\
&\quad=
\frac{
E[Y(t)e^{\eta_{W(t)}(X)}
\mathbf1\{W(t)=w,X\in B\}]
}{c_t}\\
&\quad=
P_{R_t}\{W(t)=w,X\in B\}.
\end{aligned}
\]

Thus the distribution of $(X,W(t))$
under $P_{R_t}$ is the limiting
distribution of these characteristics
among subjects experiencing an event
at time $t$. The equivalence concerns
pre-event characteristics, not the
entire observed record conditional
on failure.

\subsection{Relationship with the ordinary adoption propensity}

\paragraph{Proof of event-time overlap.}

By conditioning on adoption status,
\[
\begin{aligned}
R_1(t,x)
&=\pi_t(x)P(U\geq t\mid A<t,X=x),\\
R_0(t,x)
&=\{1-\pi_t(x)\}
P(U\geq t\mid A\geq t,X=x).
\end{aligned}
\]
Consequently,
\[
\frac{a_0(t,x)}{1-a_0(t,x)}
=
\frac{\pi_t(x)}{1-\pi_t(x)}
\rho_t(x)e^{\tau_0(x)},
\]
which establishes the logit identity
in the event-time overlap lemma.

Under Assumption~\ref{ass:overlap},
\[
\epsilon\leq\rho_t(x)
\leq\epsilon^{-1}
\]
on $[t_0,\bar t]$.
Together with
$\|\tau_0\|_\infty<\infty$, this gives
\[
\epsilon'\leq a_0(t,x)
\leq1-\epsilon',
\]
where
\[
\epsilon'=\{1+\exp(L)\}^{-1},
\qquad
L=\log\frac{1-\epsilon}{\epsilon^2}
+\|\tau_0\|_\infty.
\]
This completes the proof.
\qed
\subsection{Direct estimation from observed failures}

Equation~\eqref{eq:event-propensity-main}
suggests estimating $a_0(t,x)$
through the binary regression
\[
E\{W(U)\mid U=t,X=x,\Delta=1\}
=a_0(t,x).
\]

Within each training fold, retain one
observation per observed failure:
\[
\mathcal D_{\mathrm{event}}^{(-k)}
=
\{(U_i,X_i,W_i(U_i)):
i\in I_k^c,\Delta_i=1\}.
\]

In our simulations, we fit the
logistic generalized additive model
\begin{equation}
\label{eq:appendix-gam}
\operatorname{logit}a(t,x)
=s(t)+x^\top\alpha,
\end{equation}
where $s(t)$ is a spline in event time
with basis dimension five.

The fitted model is evaluated for all
members of each held-out risk set,
not only those who experience an event.
Although the additive specification
shares information across event times,
it may miss time--covariate interactions;
the analytical oracle below illustrates
this limitation.

Detail on grouped event times: Suppose events are recorded on intervals $I_m=[s_{m-1},s_m)$, treatment is constant within each interval, and $\Delta\Lambda_m=\Lambda_0(s_m)-\Lambda_0(s_{m-1})$. Write $u_w=e^{\eta_w(x)}\Delta\Lambda_m$ and $R_w=R_w(s_{m-1},x)$. The treated fraction among subjects failing in $I_m$ is
\[
a_m^{\mathrm{int}}(x)=\frac{R_1(1-e^{-u_1})}{R_1(1-e^{-u_1})+R_0(1-e^{-u_0})}
=a_0(s_{m-1},x)-\tfrac12\,a_0(1-a_0)(u_1-u_0)+O(u_0^2+u_1^2).
\]
The discrepancy is of the order of the per-interval event probability times $|e^{\tau_0(x)}-1|$; it vanishes when $\tau_0(x)=0$ and is negligible when events within an interval are rare. For example, a weekly event probability of 5\% and a hazard ratio of 1.5 give a centering error of about 0.003.

\subsection{Analytical oracle for the simulation DGP}
\label{sec:oracle-propensity}

For the simulation design, treatment
adoption follows
\[
A\mid X=x\sim\operatorname{Exp}\{r(x)\},
\qquad r(x)=e^{x_2+x_3},
\]
and the event hazard satisfies
\[
\lambda_0(t)=t,\qquad
\Lambda_0(t)=t^2/2,\qquad
\tau_0(x)=x_1+x_2+x_3.
\]

We calculate the two treatment-specific
risk-set probabilities analytically.
Independent censoring contributes a
common factor $G_C(t)=P(C\ge t)$,
which cancels from $a_0(t,x)$.

\paragraph{Subjects not yet treated at $t$.}

An untreated subject at time $t$ must have
$A\ge t$ and survive under the untreated
hazard until $t$. Thus
\begin{align*}
P(A\ge t\mid X=x)&=e^{-r(x)t},\\
P(T\ge t\mid A\ge t,X=x)
&=\exp\left\{
-\tfrac12e^{\eta_0(x)}t^2
\right\}.
\end{align*}
Consequently,
\begin{equation}
\label{eq:appendix-oracle-R0}
R_0(t,x)=G_C(t)
\exp\left\{
-r(x)t-\tfrac12e^{\eta_0(x)}t^2
\right\}.
\end{equation}

\paragraph{Subjects already treated at $t$.}

For an individual adopting at time $A=s<t$,
the event hazard is untreated before $s$
and treated between $s$ and $t$.
The conditional survival probability is
\[
\begin{aligned}
&P(T\ge t\mid A=s,X=x)\\
&\quad=
\exp\left\{
-\tfrac12e^{\eta_0(x)}s^2
-\tfrac12e^{\eta_1(x)}(t^2-s^2)
\right\}.
\end{aligned}
\]
Integrating over the adoption-time density
$r(x)e^{-r(x)s}$ gives
\begin{align}
R_1(t,x)
={}&G_C(t)r(x)
e^{-\frac12e^{\eta_1(x)}t^2}
\notag\\
&\times
\int_0^t
\exp\left[
-r(x)s+
\tfrac12e^{\eta_0(x)}
\{e^{\tau_0(x)}-1\}s^2
\right]\mathrm ds.
\label{eq:appendix-oracle-R1}
\end{align}

Define
\[
d(x)=\tfrac12e^{\eta_0(x)}
\{e^{\tau_0(x)}-1\},
\qquad
I(t,x)=\int_0^t
e^{-r(x)s+d(x)s^2}\,\mathrm ds.
\]
Equations~\eqref{eq:appendix-oracle-R0}
and~\eqref{eq:appendix-oracle-R1}
imply
\[
\frac{R_1(t,x)}{R_0(t,x)}
=
r(x)e^{r(x)t-d(x)t^2}I(t,x).
\]

Finally, 
\[
\operatorname{logit}a_0(t,x)
=
\log\frac{R_1(t,x)}{R_0(t,x)}
+\tau_0(x).
\]
Substituting $r(x)=e^{x_2+x_3}$ and
$\tau_0(x)=x_1+x_2+x_3$ yields
\begin{equation}
\label{eq:oracle-event-propensity}
\boxed{
\begin{aligned}
\operatorname{logit}a_0(t,x)
={}&x_1+2x_2+2x_3\\
&+r(x)t-d(x)t^2+\log I(t,x).
\end{aligned}
}
\end{equation}

The formula applies to both simulation
scenarios, with $\eta_0$ set to its
linear or nonlinear specification.
The integral $I(t,x)$ is evaluated
numerically.

The oracle formula demonstrates why
the additive GAM is generally only an
approximation: both $d(x)t^2$ and
$\log I(t,x)$ contain time--covariate
interactions absent from the additive
working model. The fitted model is therefore a working approximation, and the nuisance rate conditions of Section~\ref{sec:neyman-ortho-theory} are assumptions rather than established properties of this fit.

\section{Proofs for the DML results}
\label{sec:proof-dml}

\subsection{Technical regularity conditions}
\label{sec:dml-technical}

For a measurable scalar or vector function
$f(t,X,W(t))$, define the event-weighted norm
\begin{equation}
\label{eq:event-weighted-norm}
\|f\|_{2,\mu}^2
=
\int_0^{\bar t}
c_tE_{R_t}[\|f(t,X,W(t))\|^2]
\,d\Lambda_0(t).
\end{equation}
This integrates the squared error under
the event-risk distribution across time. Because the partial-likelihood score is unchanged when a function of time alone is added to $\nu$, nuisance errors are measured after removing such common time-only shifts.

\begin{condition}[Technical DML regularity]
\label{cond:dml-technical}
\begin{enumerate}[(i)]

\item $\beta_0$ is an interior point of
a compact convex parameter space
$\mathcal B$, and
$\Psi(\beta,a_0,\nu_0)$ has a unique
zero on $\mathcal B$.

\item The nuisance-weighted risk-set
denominators remain positive in a
neighborhood of the truth. The
expectations defining the population
score admit integrable envelopes
permitting differentiation under
expectations and time integrals.

The second derivatives
$D_{aa}^2\Psi$ and $D_{a\nu}^2\Psi$
are uniformly bounded as bilinear
operators under the norm in
Equation~\eqref{eq:event-weighted-norm},
along admissible nuisance paths.

\item The number of folds $K$ is fixed,
with $n_k/n$ bounded away from zero.
Furthermore,
\[
\sup_x\|p(x)\|<\infty,
\qquad
\sup_{x,w}|\eta_w(x)|<\infty.
\]
With probability approaching one,
the cross-fitted propensity estimates
belong to $[0,1]$, and the
cross-fitted $\hat\nu$ are uniformly
bounded.

\end{enumerate}
\end{condition}

\subsection{Proof of Proposition~\ref{prop:orthogonality}}
\label{sec:proof-prop:orthogonality}

Recall the measure $P_{R_t}$
in Equation~\eqref{eq:risk-set-measure}
and the centered regressor $q_0(t)$
defined in Section~\ref{sec:neyman-ortho-theory}.
By the definition of $a_0$,
\begin{equation}
\label{eq:proof-risk-set-centering}
E_{R_t}[q_0(t)\mid X]=0,
\qquad
E_{R_t}[q_0(t)]=0.
\end{equation}
The measure $P_{R_t}$ is held fixed
at the true event intensity throughout
the nuisance perturbations. 


Under the assumed counting-process intensity,
for any integrable predictable function $f$,
\[
E\int_0^{\bar t}f(t)\,dN(t)
=
\int_0^{\bar t}
c_t E_{R_t}[f(t)]\,d\Lambda_0(t).
\]
Thus the population score can be written as
\begin{equation}
\label{eq:proof-score-risk-measure}
\Psi(\beta,a,\nu)
=
\int_0^{\bar t}c_t
\left\{
E_{R_t}[q_a(t)]
-\bar q(t;\beta,a,\nu)
\right\}d\Lambda_0(t).
\end{equation}

We first express the risk-set mean under
$P_{R_t}$. Recall that
\[
\nu_0(t,X)+q_0(t)^\top\beta_0
=\eta_{W(t)}(X).
\]
For arbitrary $a$ and $\nu$, define
\[
H_{a,\nu}(t,X)
=
\exp\left\{
\nu(t,X)-\nu_0(t,X)
-[a(t,X)-a_0(t,X)]\tau_0(X)
\right\}.
\]

Since
\[
e^{\nu(t,X)+q_a(t)^\top\beta_0}
=
e^{\eta_{W(t)}(X)}H_{a,\nu}(t,X),
\]
the population risk-set mean satisfies
\begin{equation}
\label{eq:proof-risk-mean-change}
\bar q(t;\beta_0,a,\nu)
=
\frac{E_{R_t}[H_{a,\nu}(t,X)q_a(t)]}
{E_{R_t}[H_{a,\nu}(t,X)]}.
\end{equation}

\paragraph{Invariance to $\nu$.}

Fix $a=a_0$ and let $\nu$ be any admissible
nuisance function. Then $q_a=q_0$ and
\[
H_{a_0,\nu}(t,X)
=e^{\nu(t,X)-\nu_0(t,X)}.
\]
This factor depends only on $(t,X)$.
By Equation~\eqref{eq:proof-risk-set-centering},
\begin{align*}
&E_{R_t}[H_{a_0,\nu}(t,X)q_0(t)]\\
&\quad=
E_{R_t}\left[
H_{a_0,\nu}(t,X)
E_{R_t}[q_0(t)\mid X]
\right]
=0.
\end{align*}
Equation~\eqref{eq:proof-risk-mean-change}
therefore gives
\[
\bar q(t;\beta_0,a_0,\nu)=0.
\]
Since $E_{R_t}[q_0(t)]=0$, substituting
into Equation~\eqref{eq:proof-score-risk-measure}
yields
\[
\Psi(\beta_0,a_0,\nu)=0
\]
for every admissible $\nu$.

In particular, differentiation at $\nu_0$
gives
\[
D_\nu\Psi(\beta_0,a_0,\nu_0)[h_\nu]=0.
\]

\paragraph{Orthogonality with respect to $a$.}

Consider the joint perturbation
\[
a_s=a_0+sh_a,
\qquad
\nu_s=\nu_0+sh_\nu,
\]
where $h_a$ and $h_\nu$ are arbitrary
admissible directions and $s$ is sufficiently
close to zero.

Write
\[
q_s(t)=q_{a_s}(t)
=q_0(t)-sh_a(t,X)p(X)
\]
and
\[
H_s(t,X)
=
H_{a_s,\nu_s}(t,X)
=
\exp\left\{
s[h_\nu(t,X)-h_a(t,X)\tau_0(X)]
\right\}.
\]

First, the derivative of the observed-event
component in
Equation~\eqref{eq:proof-score-risk-measure}
is
\begin{equation}
\label{eq:proof-event-direction}
\left.
\frac{d}{ds}E_{R_t}[q_s(t)]
\right|_{s=0}
=
-E_{R_t}[h_a(t,X)p(X)].
\end{equation}

Next, by Equation~\eqref{eq:proof-risk-mean-change},
\[
\bar q(t;\beta_0,a_s,\nu_s)
=
\frac{E_{R_t}[H_s(t,X)q_s(t)]}
{E_{R_t}[H_s(t,X)]}.
\]
At $s=0$, we have $H_0=1$ and
\[
E_{R_t}[H_0q_0(t)]
=E_{R_t}[q_0(t)]=0.
\]
Differentiating the numerator gives
\begin{align*}
&\left.
\frac{d}{ds}E_{R_t}[H_s(t,X)q_s(t)]
\right|_{s=0}\\
&\quad=
-E_{R_t}[h_a(t,X)p(X)]\\
&\qquad+
E_{R_t}\left[
q_0(t)\{h_\nu(t,X)-h_a(t,X)\tau_0(X)\}
\right].
\end{align*}
The last expectation is zero because the
factor multiplying $q_0(t)$ depends only
on $(t,X)$ and
$E_{R_t}[q_0(t)\mid X]=0$.
Therefore,
\[
\left.
\frac{d}{ds}E_{R_t}[H_s q_s]
\right|_{s=0}
=
-E_{R_t}[h_a(t,X)p(X)].
\]

Since the numerator of the quotient is zero
at $s=0$ and its denominator equals one,
the quotient rule gives
\begin{equation}
\label{eq:proof-risk-mean-direction}
\left.
\frac{d}{ds}
\bar q(t;\beta_0,a_s,\nu_s)
\right|_{s=0}
=
-E_{R_t}[h_a(t,X)p(X)].
\end{equation}

Equations~\eqref{eq:proof-event-direction}
and~\eqref{eq:proof-risk-mean-direction}
are identical. Their contributions therefore
cancel in the population score:
\[
\left.
\frac{d}{ds}
\Psi(\beta_0,a_s,\nu_s)
\right|_{s=0}
=0.
\]
Taking $h_\nu=0$ gives
\[
D_a\Psi(\beta_0,a_0,\nu_0)[h_a]=0.
\]
Together with the result for $\nu$, this
establishes Neyman orthogonality with
respect to both nuisance functions.
\qed

\subsection{Proof of Proposition~\ref{prop:dml-rate}}
\label{sec:proof-dml-rate}

Write $g=(a,\nu)$ and
$g_0=(a_0,\nu_0)$.
For each fold $k$, let
$\hat g_k=(\hat a^{(-k)},\hat\nu^{(-k)})$
and $\kappa_k=n_k/n$.

\paragraph{Step 1: Population nuisance remainder.}

Let $\delta a=a-a_0$,
$\delta\nu=\nu-\nu_0$, and
$H=H_{a,\nu}$ as defined in the
proof of Proposition~\ref{prop:orthogonality}.

Using
$q_a=q_0-\delta a\,p$ and
$E_{R_t}[q_0(t)\mid X]=0$,
Equation~\eqref{eq:proof-score-risk-measure}
gives the exact identity
\begin{equation}
\label{eq:proof-exact-covariance}
\Psi(\beta_0,a,\nu)
=
\int_0^{\bar t}c_t
\frac{
\operatorname{Cov}_{R_t}
\{\delta a\,p,H\}
}{
E_{R_t}[H]
}
\,d\Lambda_0(t),
\end{equation}
where covariances are evaluated at
time $t$ and interpreted componentwise.

At the truth, $H=1$ and
\[
DH[h_a,h_\nu]=h_\nu-h_a\tau_0.
\]
Differentiating
Equation~\eqref{eq:proof-exact-covariance}
therefore yields
\begin{align}
&D_{aa}^2\Psi[h_a,k_a]
\nonumber\\
&\quad=
-\int_0^{\bar t}c_t
\left[
\operatorname{Cov}_{R_t}
(h_ap,k_a\tau_0)
+
\operatorname{Cov}_{R_t}
(k_ap,h_a\tau_0)
\right]d\Lambda_0(t),
\label{eq:proof-second-aa}\\
&D_{a\nu}^2\Psi[h_a,h_\nu]
=
\int_0^{\bar t}c_t
\operatorname{Cov}_{R_t}
(h_ap,h_\nu)\,d\Lambda_0(t),
\label{eq:proof-second-anu}\\
&D_{\nu\nu}^2\Psi[h_\nu,k_\nu]=0.
\label{eq:proof-second-nunu}
\end{align}
All derivatives are evaluated at
$(\beta_0,g_0)$.

By Proposition~\ref{prop:orthogonality},
$\Psi(\beta_0,a_0,\nu)=0$
for every admissible $\nu$.
Applying the fundamental theorem of
calculus first along the $a$ direction
and then to its first derivative gives
\begin{align*}
&\Psi(\beta_0,a,\nu)-\Psi(\beta_0,g_0)\\
&=
\int_0^1\int_0^s
D_{aa}^2\Psi(
\beta_0,a_0+u\delta a,\nu)
[\delta a,\delta a]\,du\,ds\\
&\quad+
\int_0^1
D_{a\nu}^2\Psi(
\beta_0,a_0,\nu_0+v\delta\nu)
[\delta a,\delta\nu]\,dv.
\end{align*}

The bilinear derivative bounds in
Condition~\ref{cond:dml-technical}(ii)
and Cauchy--Schwarz imply
\[
\|\Psi(\beta_0,a,\nu)
-\Psi(\beta_0,g_0)\|
\leq C\left(
\|\delta a\|_{2,\mu}^2+
\|\delta a\|_{2,\mu}
\|\delta\nu\|_{2,\mu}
\right).
\]

Applying this bound fold by fold gives
\begin{equation}
\label{eq:nuisance-remainder}
B_n:=
\sum_{k=1}^K\kappa_k
\Psi(\beta_0,\hat g_k)
=
O_p\{r_{a,n}^2+r_{a,n}r_{\nu,n}\}.
\end{equation}

\paragraph{Step 2: Oracle empirical score.}

For any function $f$, write
\[
P_{n,k}f=\frac1{n_k}
\sum_{i\in I_k}f(O_i).
\]
Define the oracle empirical risk-set
expectation
\[
\langle f\rangle_{n,k,t}
=
\frac{
P_{n,k}[Z_t f(t)]
}{
P_{n,k}Z_t
},
\qquad
Z_t=Y(t)e^{\eta_{W(t)}(X)}.
\]
Let
\[
\bar q_{n,k}^0(t)
=
\langle q_0(t)\rangle_{n,k,t}.
\]

The oracle fold score is
\[
\Psi_{n,k}^0
=
P_{n,k}\int_0^{\bar t}
\{q_0(t)-\bar q_{n,k}^0(t)\}
\,dN(t).
\]
Using the Doob--Meyer decomposition
\[
dN_i(t)=Z_{i,t}\,d\Lambda_0(t)
+dM_i(t),
\]
we obtain
\begin{align*}
\Psi_{n,k}^0
={}&P_{n,k}\int_0^{\bar t}
\{q_0(t)-\bar q_{n,k}^0(t)\}
\,dM(t)\\
&+
\int_0^{\bar t}
P_{n,k}\left[
Z_t\{q_0(t)-\bar q_{n,k}^0(t)\}
\right]d\Lambda_0(t).
\end{align*}

The second term is exactly zero because
\[
P_{n,k}[Z_tq_0(t)]
=
P_{n,k}Z_t\,\bar q_{n,k}^0(t).
\]
Thus
\begin{equation}
\label{eq:proof-oracle-martingale}
\Psi_{n,k}^0
=
P_{n,k}\int_0^{\bar t}q_0(t)\,dM(t)
-
\int_0^{\bar t}
\bar q_{n,k}^0(t)\,P_{n,k}dM(t).
\end{equation}

The population conditional centering identity
gives $E[Z_tq_0(t)]=0$.
Independence across subjects and the
bounded-envelope conditions imply
\[
E\int_0^{\bar t}
\|P_{n,k}[Z_tq_0(t)]\|^2
\,d\Lambda_0(t)
=
O(n_k^{-1}).
\]
Since $P_{n,k}Z_t$ is uniformly bounded
away from zero with probability
approaching one,
\[
\int_0^{\bar t}
\|\bar q_{n,k}^0(t)\|^2
\,d\Lambda_0(t)
=
O_p(n_k^{-1}).
\]

The empirical risk-set mean is
predictable. Applying the martingale
isometry, with localization on the
event that risk-set denominators
are bounded away from zero, gives
\[
\int_0^{\bar t}
\bar q_{n,k}^0(t)\,P_{n,k}dM(t)
=
O_p(n_k^{-1}).
\]
Therefore,
\[
\Psi_{n,k}^0
=
\frac1{n_k}\sum_{i\in I_k}\psi_i
+
O_p(n_k^{-1}),
\]
where we recall in main text
\[
\psi_i=\int_0^{\bar t}q_i^0(t)\,dM_i(t).
\]

Aggregating over the fixed number
of folds yields
\begin{equation}
\label{eq:proof-oracle-linear}
\Psi_n(\beta_0,g_0)
=
\frac1n\sum_{i=1}^n\psi_i
+O_p(n^{-1})
=
O_p(n^{-1/2}).
\end{equation}

\paragraph{Step 3: Cross-fitting remainder.}

We now account for nuisance estimation
and the empirical risk-set ratios.

For each fold, let
\[
\delta a_k=\hat a^{(-k)}-a_0,
\qquad
\delta\nu_k=\hat\nu^{(-k)}-\nu_0,
\]
and define
\[
H_k(t,X)
=
e^{\delta\nu_k(t,X)
-\delta a_k(t,X)\tau_0(X)}.
\]
Put
\[
\rho_{k,n}
=
\|\delta a_k\|_{2,\mu}
+
\|\delta\nu_k\|_{2,\mu}.
\]

Conditional on the training observations
$I_k^c$, these functions are fixed and
the held-out observations remain
independent.

Write
$\widehat\Psi_{n,k}(\beta,g)$
for the fold-specific empirical Cox score
computed with nuisance functions $g$.
Define
\begin{align}
\mathcal E_n
={}&
\sum_{k=1}^K\kappa_k
\Big[
\widehat\Psi_{n,k}(\beta_0,\hat g_k)
-\Psi(\beta_0,\hat g_k)
\nonumber\\
&\hspace{27mm}
-\widehat\Psi_{n,k}(\beta_0,g_0)
+\Psi(\beta_0,g_0)
\Big].
\label{eq:proof-crossfit-remainder}
\end{align}

We establish the order of this term.
Under the true intensity, the empirical
score can be decomposed into a
martingale component and a compensator
component. For an arbitrary $g=(a,\nu)$,
the latter is
\[
D_{n,k}(g)
=
\int_0^{\bar t}
P_{n,k}Z_t
\left\{
\langle q_a(t)\rangle_{n,k,t}
-\bar q_{n,k}(t;\beta_0,g)
\right\}d\Lambda_0(t).
\]

At $g=\hat g_k$, the same change-of-measure
calculation as in Step 1 gives
\begin{align}
D_{n,k}(\hat g_k)
={}&
\int_0^{\bar t}
\frac{P_{n,k}Z_t}
{\langle H_k\rangle_{n,k,t}}
\Big[
\operatorname{Cov}_{n,k,t}
\{\delta a_k p,H_k\}
\nonumber\\
&\hspace{27mm}
-
\operatorname{Cov}_{n,k,t}
\{q_0,H_k\}
\Big]d\Lambda_0(t),
\label{eq:proof-empirical-drift}
\end{align}
where $\operatorname{Cov}_{n,k,t}$
denotes covariance under the empirical
weighted risk-set distribution
$\langle\cdot\rangle_{n,k,t}$.

At $g=g_0$, $H_k=1$, and therefore
\[
D_{n,k}(g_0)=0
\]
exactly.

By contrast, the population drift is
\[
\Psi(\beta_0,\hat g_k)
=
\int_0^{\bar t}
c_t
\frac{
\operatorname{Cov}_{R_t}
\{\delta a_k p,H_k\}
}{
E_{R_t}[H_k]
}
\,d\Lambda_0(t),
\]
because
$\operatorname{Cov}_{R_t}(q_0,H_k)=0$.

To compare these expressions, note that
bounded predictors and the mean-value
theorem give
\[
|H_k(t,X)-1|
\leq
C\{
|\delta\nu_k(t,X)|
+
|\delta a_k(t,X)|
\}.
\]
Consequently,
\[
\|H_k-1\|_{2,\mu}
\leq C\rho_{k,n}.
\]

In particular, conditional on the
training sample, the centered
empirical averages of $Z_tq_0(t)
\{H_k(t,X)-1\}$ have integrated
second moments bounded by
$C n_k^{-1}\rho_{k,n}^2$.
Their population expectations vanish
by $E_{R_t}[q_0(t)\mid X]=0$.

The remaining terms in
Equation~\eqref{eq:proof-empirical-drift}
are products and ratios of empirical
weighted averages. Their denominators
are bounded away from zero, and
their numerators satisfy the same
integrated second-moment bounds.
Expanding the ratios around their
population counterparts and applying
Cauchy--Schwarz gives
\begin{equation}
\label{eq:proof-drift-bound}
D_{n,k}(\hat g_k)
-\Psi(\beta_0,\hat g_k)
=
O_p\left(
n_k^{-1/2}\rho_{k,n}+n_k^{-1}
\right).
\end{equation}

For the martingale component, the
difference between the fitted and
oracle score integrands is
\[
-\delta a_k(t,X)p(X)
-
\left\{
\bar q_{n,k}(t;\beta_0,\hat g_k)
-\bar q_{n,k}^0(t)
\right\}.
\]
The bounded-envelope and denominator
conditions, together with the
$L_2(\mu)$ nuisance rates, imply that
its integrated squared intensity norm
is $O_p(\rho_{k,n}^2)$.
The martingale isometry therefore
bounds the difference between the
fitted and oracle martingale
components by
\[
O_p(n_k^{-1/2}\rho_{k,n}).
\]

Combining this bound with
Equation~\eqref{eq:proof-drift-bound}
and summing over the fixed number
of folds yields
\begin{equation}
\label{eq:proof-crossfit-bound}
\mathcal E_n
=
O_p\left\{
n^{-1/2}(r_{a,n}+r_{\nu,n})
+n^{-1}
\right\}
=o_p(n^{-1/2}).
\end{equation}
This establishes the stochastic
remainder bound using the actual
fold-restricted Cox risk sets,
without imposing a Donsker
condition on the nuisance learners.

\paragraph{Step 4: Consistency and rate.}

The bounded-envelope conditions and
fold-conditional laws of large numbers
give uniform convergence of the
cross-fitted score:
\[
\sup_{\beta\in\mathcal B}
\left\|
\widehat\Psi_n(\beta,\hat g)
-\Psi(\beta,g_0)
\right\|
=o_p(1).
\]
Here $\widehat\Psi_n(\beta,\hat g)$
denotes the aggregate fold-specific
score, rather than a score evaluated
at one common nuisance estimate.

By compactness of $\mathcal B$,
continuity of the population score,
and uniqueness of its zero at
$\beta_0$, any solution
$\hat\beta\in\mathcal B$ to the
cross-fitted estimating equation
satisfies
\[
\hat\beta\xrightarrow{p}\beta_0.
\]

Differentiating the population score
with respect to $\beta$ at the truth
gives
\begin{align*}
-\partial_\beta\Psi(\beta_0,g_0)
&=
\int_0^{\bar t}
c_t\operatorname{Var}_{R_t}
\{q_0(t)\}\,d\Lambda_0(t)\\
&=
\int_0^{\bar t}
c_tE_{R_t}[q_0(t)q_0(t)^\top]
\,d\Lambda_0(t)
=J,
\end{align*}
where the second equality uses
$E_{R_t}[q_0(t)]=0$.

The same fold-conditional laws of
large numbers, applied to the
risk-set second moments, give local
uniform convergence of the empirical
score derivative to its population
counterpart. Continuity of the latter,
together with $\hat\beta\to_p\beta_0$,
therefore implies
\[
-\int_0^1
\partial_\beta\widehat\Psi_n
\{\beta_0+s(\hat\beta-\beta_0),\hat g\}
\,ds
=
J+o_p(1).
\]

Applying the fundamental theorem
of calculus to the estimating
equation yields
\begin{align*}
0
={}&
\widehat\Psi_n(\beta_0,\hat g)
-\{J+o_p(1)\}(\hat\beta-\beta_0)\\
={}&
\Psi_n(\beta_0,g_0)
+B_n+\mathcal E_n\\
&-\{J+o_p(1)\}(\hat\beta-\beta_0).
\end{align*}

Since $J$ is nonsingular,
Equations~\eqref{eq:nuisance-remainder},
\eqref{eq:proof-oracle-linear}, and
\eqref{eq:proof-crossfit-bound} imply
\[
\|\hat\beta-\beta_0\|
=
O_p\left\{
n^{-1/2}
+r_{a,n}^2
+r_{a,n}r_{\nu,n}
\right\}.
\]
This proves
Proposition~\ref{prop:dml-rate}.
\qed

\subsection{Proof of Proposition~\ref{prop:dml-clt}}
\label{sec:proof-dml-clt}

Under the stated rate condition,
\[
\sqrt n\,r_{a,n}(r_{a,n}+r_{\nu,n})
\longrightarrow 0,
\]
Equation~\eqref{eq:nuisance-remainder}
implies
\[
B_n=o_p(n^{-1/2}).
\]
Moreover,
Equation~\eqref{eq:proof-crossfit-bound}
gives
\[
\mathcal E_n=o_p(n^{-1/2}).
\]

The score expansion established in the
proof of Proposition~\ref{prop:dml-rate}
therefore reduces to
\[
0=
\Psi_n(\beta_0,g_0)
-\{J+o_p(1)\}(\hat\beta-\beta_0)
+o_p(n^{-1/2}).
\]

Since $J$ is nonsingular,
\[
\sqrt n(\hat\beta-\beta_0)
=
J^{-1}\sqrt n\,
\Psi_n(\beta_0,g_0)+o_p(1).
\]
By Equation~\eqref{eq:proof-oracle-linear},
\begin{equation}
\label{eq:dml-linearization}
\sqrt n(\hat\beta-\beta_0)
=
J^{-1}\frac1{\sqrt n}
\sum_{i=1}^n\psi_i+o_p(1),
\end{equation}
where
\[
\psi_i
=
\int_0^{\bar t}q_i^0(t)\,dM_i(t).
\]

Under the correctly specified
counting-process intensity,
each $M_i$ is a square-integrable
martingale. Since $q_i^0$ is
predictable,
\[
E[\psi_i]=0.
\]

The martingale isometry gives
\begin{align*}
E[\psi_i\psi_i^\top]
&=
E\int_0^{\bar t}
q_i^0(t)q_i^0(t)^\top
\,d\langle M_i\rangle(t)\\
&=
E\int_0^{\bar t}
Y_i(t)e^{\eta_{W_i(t)}(X_i)}
q_i^0(t)q_i^0(t)^\top
\,d\Lambda_0(t)\\
&=J.
\end{align*}

In particular, $\Sigma=E[\psi_i\psi_i^\top]$
is finite and positive definite under
Condition~\ref{cond:dml}(i).
The vectors $\psi_i$ are independent
and identically distributed across
subjects. Thus the multivariate
central limit theorem yields
\[
\frac1{\sqrt n}\sum_{i=1}^n\psi_i
\rightsquigarrow N(0,\Sigma).
\]

Combining this convergence with
Equation~\eqref{eq:dml-linearization}
and applying Slutsky's theorem gives
\[
\sqrt n(\hat\beta-\beta_0)
\rightsquigarrow
N(0,J^{-1}\Sigma J^{-1}).
\]
This proves
Proposition~\ref{prop:dml-clt}.
\qed

\paragraph{Variance estimation.}

We briefly verify consistency of the
plug-in covariance estimator in
Section~\ref{sec:DML}.

At the truth, the population risk-set
mean of $q_0(t)$ vanishes. Hence,
by Equation~\eqref{eq:information-J},
the fold-conditional laws of large
numbers and nuisance consistency give
\[
\hat J\xrightarrow{p}
\int_0^{\bar t}
c_tE_{R_t}[q_0(t)q_0(t)^\top]
\,d\Lambda_0(t)
=J.
\]

For each fold, the Breslow estimator
satisfies the exact decomposition
\[
d\hat\Lambda_{0,k}(t)
=
\frac{P_{n,k}dM(t)}
{\hat S_k^{(0)}(t;\hat\beta)}
+
\frac{P_{n,k}Z_t}
{\hat S_k^{(0)}(t;\hat\beta)}
\,d\Lambda_0(t).
\]
Under the bounded-envelope and
risk-set stability conditions,
nuisance consistency and
$\hat\beta\to_p\beta_0$ imply
\[
\frac{P_{n,k}Z_t}
{\hat S_k^{(0)}(t;\hat\beta)}
\longrightarrow 1
\]
in the integrated risk-set norm.
The martingale term in the Breslow
decomposition is $O_p(n_k^{-1/2})$
when integrated against uniformly
bounded predictable score integrands.

Using this decomposition and the
$L_2(\mu)$ convergence of
$\hat q_i$ to $q_i^0$, the estimated
subject-level influence scores satisfy
\[
\frac1n\sum_{i=1}^n
\|\hat\psi_i-\psi_i\|^2=o_p(1).
\]
Here replacing $\beta_0$ by
$\hat\beta$ is justified by
finite-dimensional local
stochastic equicontinuity.

Consequently, Cauchy--Schwarz
and the law of large numbers yield
\[
\hat\Sigma
=
\frac1n\sum_{i=1}^n
\hat\psi_i\hat\psi_i^\top
\xrightarrow{p}
E[\psi_i\psi_i^\top]
=\Sigma.
\]
Since $J$ is nonsingular,
the continuous mapping theorem gives
\[
\hat\Omega
=
\hat J^{-1}\hat\Sigma\hat J^{-1}
\xrightarrow{p}
J^{-1}\Sigma J^{-1}
=\Omega.
\]

\section{Additional Simulation Results}
\subsection{Complete simulation tables}
\label{sec:simulation-tables}

See Tables~\ref{tab:sim-linear-full} to \ref{tab:cov-nonlinear-full}. They are listed for completeness. Key observations are discussed in the main paper. 


\begin{table}[h]
\centering
\caption{MSE under linear $\eta_0$. All $\eta_0$ and propensity specifications. Lower is better.}
\label{tab:sim-linear-full}
\begin{tabular}{llcccc}
\toprule
Method & $\eta_0$ / prop.\ spec & $n=200$ & $n=500$ & $n=1000$ & $n=2000$ \\
\midrule
S-Cox & complex & 0.423 & 0.124 & 0.056 & 0.028 \\
S-Cox & linear & 0.230 & 0.078 & 0.034 & 0.018 \\
\midrule
S-Lasso & complex & 0.298 & 0.106 & 0.049 & 0.025 \\
S-Lasso & linear & 0.230 & 0.078 & 0.034 & 0.018 \\
\midrule
TV-CSL & complex / intercept-only & 1.448 & 0.730 & 0.447 & 0.294 \\
TV-CSL & complex / marg.\ prop. & 1.963 & 0.676 & 0.349 & 0.193 \\
TV-CSL & complex / time-var.\ prop. & 2.435 & 0.510 & 0.218 & 0.103 \\
TV-CSL & linear / intercept-only & 0.398 & 0.110 & 0.045 & 0.022 \\
TV-CSL & linear / marg.\ prop. & 0.694 & 0.168 & 0.066 & 0.031 \\
TV-CSL & linear / time-var.\ prop. & 0.881 & 0.164 & 0.065 & 0.030 \\
\bottomrule
\end{tabular}
\end{table}


\begin{table}[h]
\centering
\caption{MSE under non-linear $\eta_0$. All $\eta_0$ and propensity specifications.}
\label{tab:sim-nonlinear-full}
\begin{tabular}{llcccc}
\toprule
Method & $\eta_0$ / prop.\ spec & $n=200$ & $n=500$ & $n=1000$ & $n=2000$ \\
\midrule
S-Cox & complex & 0.377 & 0.164 & 0.112 & 0.095 \\
S-Cox & linear & 0.661 & 0.597 & 0.570 & 0.565 \\
\midrule
S-Lasso & complex & 0.703 & 0.370 & 0.221 & 0.164 \\
S-Lasso & linear & 0.661 & 0.597 & 0.570 & 0.565 \\
\midrule
TV-CSL & complex / intercept-only & 0.406 & 0.219 & 0.129 & 0.078 \\
TV-CSL & complex / marg.\ prop. & 0.895 & 0.229 & 0.095 & 0.041 \\
TV-CSL & complex / time-var.\ prop. & 2.134 & 0.282 & 0.098 & 0.043 \\
TV-CSL & linear / intercept-only & 0.563 & 0.462 & 0.447 & 0.444 \\
TV-CSL & linear / marg.\ prop. & 0.742 & 0.188 & 0.105 & 0.080 \\
TV-CSL & linear / time-var.\ prop. & 1.825 & 0.240 & 0.092 & 0.050 \\
\bottomrule
\end{tabular}
\end{table}


\begin{table}[h]
\centering
\caption{Empirical 95\% CI coverage under linear $\eta_0$ (linear $\eta_0$ spec). Averaged over $\beta_2,\beta_3,\beta_4$.}
\label{tab:cov-linear-full}
\begin{tabular}{llcccc}
\toprule
Method & prop.\ spec & $n=200$ & $n=500$ & $n=1000$ & $n=2000$ \\
\midrule
S-Cox & --- & 0.933 & 0.945 & 0.956 & 0.942 \\
\midrule
TV-CSL (information)    & intercept-only & 0.901 & 0.892 & 0.902 & 0.878 \\
TV-CSL (sandwich) & intercept-only & 0.754 & 0.824 & 0.863 & 0.857 \\
TV-CSL (information)    & marg.\ prop. & 0.947 & 0.947 & 0.953 & 0.948 \\
TV-CSL (sandwich) & marg.\ prop. & 0.773 & 0.874 & 0.918 & 0.927 \\
TV-CSL (information)    & time-var.\ prop. & 0.928 & 0.948 & 0.948 & 0.947 \\
TV-CSL (sandwich) & time-var.\ prop. & 0.721 & 0.864 & 0.907 & 0.923 \\
\bottomrule
\end{tabular}
\end{table}


\begin{table}[h]
\centering
\caption{Empirical 95\% CI coverage under non-linear $\eta_0$ (complex $\eta_0$ spec). Averaged over $\beta_2,\beta_3,\beta_4$.}
\label{tab:cov-nonlinear-full}
\begin{tabular}{llcccc}
\toprule
Method & prop.\ spec & $n=200$ & $n=500$ & $n=1000$ & $n=2000$ \\
\midrule
S-Cox & --- & 0.920 & 0.851 & 0.749 & 0.540 \\
\midrule
TV-CSL (information)    & intercept-only & 0.713 & 0.524 & 0.476 & 0.383 \\
TV-CSL (sandwich) & intercept-only & 0.605 & 0.498 & 0.475 & 0.391 \\
TV-CSL (information)    & marg.\ prop. & 0.896 & 0.912 & 0.921 & 0.929 \\
TV-CSL (sandwich) & marg.\ prop. & 0.753 & 0.861 & 0.906 & 0.924 \\
TV-CSL (information)    & time-var.\ prop. & 0.855 & 0.922 & 0.927 & 0.934 \\
TV-CSL (sandwich) & time-var.\ prop. & 0.619 & 0.851 & 0.901 & 0.921 \\
\bottomrule
\end{tabular}
\end{table}

\subsection{Isolating the Effect of the Propensity Centering}\label{sec:centering-isolation}

In the main simulation, marginal and time-varying centering perform similarly under $\eta_0^{NL}$. This does not mean that the centering is unimportant. By Proposition~\ref{prop:dml-rate}, the bias of TV-CSL is of the order of the product of the propensity error and the outcome-model error. If the outcome learner is nearly exact, any reasonable centering gives small bias. If the outcome learner is far from $\eta_0$, the exact $a_0(t,x)$ still removes the population bias, but whether a better estimated centering helps depends on how the two nuisance errors co-vary. Under $\eta_0^{NL}$ the complex specification is in the second case: its fitted $\hat\eta_0$ has root-mean-squared error equal to $0.64$ of the standard deviation of $\eta_0^{NL}$. At $n=4000$ and $\beta_0=(0,2,2,2)^\top$, replacing $\hat\eta_0$ and $\hat\eta_1$ with the true functions reduces the slope bias of every centering to at most $0.014$ in absolute value, with coverage between $0.93$ and $0.95$. With the estimated functions, the slope bias is $-0.102$ for marginal centering, $-0.129$ for time-varying centering, and $-0.007$ for the oracle.

To compare the centerings, we therefore need a baseline risk for which the outcome-model error is neither negligible nor dominant. We use the family
\[
\eta_0^{(w)}(x)=w\,\eta_0^{NL}(x)+(1-w)\,\eta_0^{S}(x),
\qquad
\eta_0^{S}(x)=\tfrac34\{\sin x_1+x_2^2/2+x_1x_3/2\},
\qquad w\in[0,1].
\]
The function $\eta_0^{S}$ is non-linear and non-additive, but the complex specification recovers it well. The weight $w$ moves the baseline risk from $\eta_0^{S}$ ($w=0$) to $\eta_0^{NL}$ ($w=1$).

We choose $w$ using only the first-stage outcome learner, before fitting any treatment-effect estimator. For each $w\in\{0,0.25,0.5,0.75,1\}$, we fit the complex specification at $n=4000$ and compute the root-mean-squared error of $\hat\eta_0$ relative to the standard deviation of $\eta_0^{(w)}$, after centering both at their means. We fixed the target interval $[0.2,0.4]$ in advance. Table~\ref{tab:w-grid} reports the relative errors. Only $w=0.25$ falls in the interval.

\begin{table}[ht]
\centering
\caption{Relative error of the complex outcome fit, $\{\text{RMSE of }\hat\eta_0\}/\operatorname{sd}(\eta_0)$, at $n=4000$ and $\beta_0=(0,2,2,2)^\top$, averaged over two datasets.}
\label{tab:w-grid}
\begin{tabular}{lccccc}
\hline
$w$ & 0 & 0.25 & 0.5 & 0.75 & 1\\
\hline
Relative error & 0.11 & 0.28 & 0.58 & 0.70 & 0.64\\
\hline
\end{tabular}
\end{table}

The design is otherwise that of Section~\ref{sec:simulation-design}, with $\eta_0=\eta_0^{(0.25)}$, $n=4000$, and $\beta_0=(0,m,m,m)^\top$ for $m\in\{0,0.5,1,1.5,2\}$. Each value of $m$ uses the same 50 simulated datasets up to the event times, so differences across $m$ are not driven by different covariate draws. We fit TV-CSL with the complex outcome specification and the marginal and time-varying propensities of Section~\ref{sec:methods-comparison}, together with the oracle propensity of Appendix~\ref{sec:oracle-propensity}, which uses the true $a_0(t,x)$. The case $m=0$ is a null check. When $\tau_0=0$, treatment does not change the hazard, so $a_0(t,x)=P(A<t\mid X=x)$ and the marginal and event propensities share the same target $\pi_t(x)$, although the fitted working models need not recover it exactly.

\begin{table}[ht]
\centering
\caption{TV-CSL under $\eta_0^{(0.25)}$ at $n=4000$, by treatment-effect scale $m$. Bias is averaged over the three slope coefficients. Coverage entries are information/sandwich. Columns are marginal, time-varying, and oracle propensity.}
\label{tab:centering-sweep}
\begin{tabular}{c ccc ccc ccc}
\hline
 & \multicolumn{3}{c}{Bias} & \multicolumn{3}{c}{Coverage} & \multicolumn{3}{c}{MSE}\\
$m$ & marg. & t.v. & oracle & marg. & t.v. & oracle & marg. & t.v. & oracle\\
\hline
0   & $-0.004$ & $-0.006$ & $-0.004$ & 0.88/0.87 & 0.88/0.87 & 0.88/0.87 & 0.013 & 0.013 & 0.013\\
0.5 & $-0.007$ & $-0.009$ & $-0.003$ & 0.93/0.91 & 0.91/0.90 & 0.93/0.92 & 0.013 & 0.014 & 0.013\\
1   & $-0.018$ & $-0.012$ & $0.001$  & 0.91/0.89 & 0.91/0.89 & 0.93/0.91 & 0.015 & 0.016 & 0.015\\
1.5 & $-0.033$ & $-0.022$ & $0.000$  & 0.89/0.88 & 0.91/0.90 & 0.93/0.93 & 0.018 & 0.019 & 0.018\\
2   & $-0.048$ & $-0.031$ & $-0.002$ & 0.83/0.80 & 0.90/0.87 & 0.91/0.91 & 0.025 & 0.024 & 0.022\\
\hline
\end{tabular}
\end{table}

Table~\ref{tab:centering-sweep} reports the results. At $m=0$ the three centerings are indistinguishable, as expected. As $m$ grows, the bias of marginal centering grows, from $-0.004$ at $m=0$ to $-0.048$ at $m=2$, and its information-based coverage falls to $0.83$. Time-varying centering reduces this bias by about one third at every $m\geq1$. Paired over datasets, the reduction is $0.006$, $0.011$, and $0.017$ at $m=1$, $1.5$, and $2$, with $z$-statistics $2.7$, $4.2$, and $4.9$. At $m=2$, time-varying centering keeps coverage at $0.90$. The oracle is unbiased at every $m$ and shows how much of the remaining bias is due to error in the estimated $a_0$.

Time-varying centering does not clearly reduce MSE: the differences are at most $0.001$, within Monte Carlo error. Its estimates are more variable than those of marginal centering, with empirical standard deviation $0.078$ against $0.071$ at $m=2$, and this offsets the smaller bias. The gain from the correct centering is therefore in bias and coverage, which is what Proposition~\ref{prop:dml-rate} concerns. No centering reaches $0.95$, including the oracle at $m=0$, so the remaining undercoverage is not due to the centering; it may reflect variance estimation, non-normality, outcome-model error, or Monte Carlo noise (with 50 replications, the Monte Carlo standard error of a coverage near $0.90$ is about $0.04$).

For comparison, the same experiment with $w=1$, that is, under $\eta_0^{NL}$, gives similar bias for the two estimated centerings at every $m$; while the oracle remains unbiased. This is the case in which the outcome-model error dominates.

\section{Data Analysis: Sensitivity Table}
See Table~\ref{tab:copilot-sensitivity}.

\input{tables/tab_copilot_sensitivity}

\section{Additional Application: Stanford Heart Transplant Data}
\label{sec:stanford-appendix}

The Stanford Heart Transplant data come from the Stanford transplantation programme, which began in the 1960s. Patients entered the study with severe heart disease and waited for a compatible donor heart. The data follow survival from programme acceptance through the waiting and post-transplant periods and include age, prior surgery, and year of acceptance \citep{crowley1977covariance}. We use this small dataset as an additional illustration of the importance of updating treatment status at transplant time. In light of the simulation results and the sample size of 103, the direct time-dependent Cox analysis is the primary analysis; the flexible comparison below is treated as a finite-sample diagnostic.

\subsection{Summary statistics}

\begin{figure}[ht]
\centering
\includegraphics[width=0.8\linewidth]{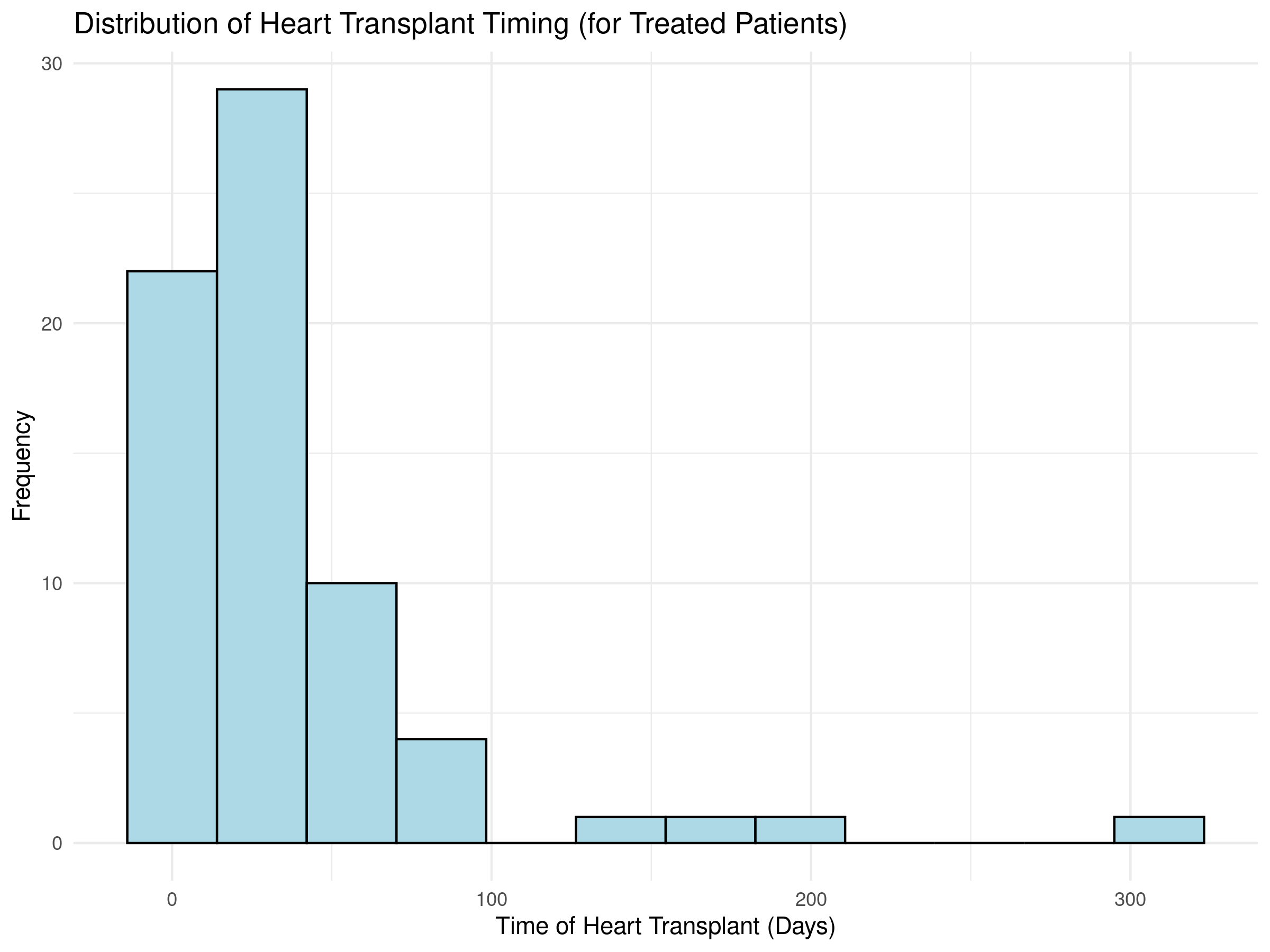}
\caption{Distribution of Time to Heart Transplant}
\label{fig:time_variation_stanford-rhc}
\end{figure}

The regressors are age at acceptance, an indicator of prior surgery, and year of acceptance measured from the start of the programme in 1967. We use time since programme acceptance as the survival time and include year of acceptance as a baseline covariate. This keeps time zero clinically comparable across patients while allowing outcomes to change over calendar time. Table~\ref{tab:summary-stat-stanford-rhc} gives the summary statistics. The average age is 45.17 years, 16\% of patients had prior surgery, and 69 of the 103 patients received a transplant. Figure~\ref{fig:time_variation_stanford-rhc} shows the variation in transplant time.

\begin{table}[ht]
\centering
\caption{Summary Statistics for the Stanford Heart Transplant Data}
\label{tab:summary-stat-stanford-rhc}
\begin{tabular}{lrr}
\toprule
Variable & Mean & SD \\
\midrule
Age & 45.17 & 9.80 \\
Prior surgery & 0.16 & 0.36 \\
Year of acceptance & 3.36 & 1.86 \\
Received transplant & 0.67 & 0.47 \\
\bottomrule
\end{tabular}
\end{table}

\subsection{Effect of Ignoring Transplant Time}

In this section, we compare two Cox models using all 103 patients. The first treats eventual transplant receipt as a fixed treatment variable: a patient is classified as treated if they receive a transplant at any point during follow-up, regardless of when it occurs. The second uses the time-varying treatment process in Equation~\ref{eq:ordinary-cox-hazard}, with transplant as adoption, so that each patient contributes untreated time while waiting and treated time after transplant. This can be seen as a replication of \cite{crowley1977covariance}.

When transplant and death or censoring occur on the same day, the transplant is recorded half a day earlier, so the event counts as treated; this affects one patient. Both models adjust for age, year of acceptance, and prior surgery. We use age and year of acceptance, both standardized, as treatment-effect modifiers. Prior surgery is an adjustment covariate only, because just 16 patients had prior surgery. As in Section~\ref{sec:copilot-application}, we report hazard ratios. Here, a hazard ratio of \(h\) means that patients who had received a transplant were approximately \(h\) times as likely, at any given time, to die as comparable patients who were still waiting.

Table~\ref{tab:impact-time-varying-HTE} shows that the two models give very different answers. When transplant time is ignored, transplant appears strongly protective: for a patient of average age accepted in an average year, the hazard ratio is 0.19 (log hazard ratio $-1.648$, $p < 0.001$). When transplant status is updated at the transplant date, the hazard ratio is 1.02 (log hazard ratio $0.020$, $p = 0.951$), and there is no evidence that transplant changed mortality for this patient. The interactions also change: the age interaction changes sign, from $-0.228$ to $0.296$, and the year interaction increases from $0.143$ to $0.394$, although neither is significant in either model. The fixed model counts the waiting period of eventual recipients as treated time. Because a patient must survive the wait to receive a transplant, this period is event-free by construction, and the fixed model credits it to transplant. The difference between the two models is therefore consistent with immortal-time bias. The semi-synthetic study in the next section confirms this.

\input{tables/tab_stanford_ignore_time}

\subsection{Semi-Synthetic Comparison at the Stanford Sample Size}

In this section, we use a semi-synthetic study to answer two questions that the real data cannot answer on their own. First, is the difference in Table~\ref{tab:impact-time-varying-HTE} due to immortal-time bias? Second, with only 103 patients, does TV-CSL improve on the time-varying Cox model? The true effects are unknown in the real data, so we construct data sets in which they are known but which otherwise resemble the Stanford data.

We keep each patient's real covariates and real censoring date, and redraw transplant and death times. Death times follow the time-varying Cox model fitted to the real data in the previous section, so the true effects are the time-varying estimates in Table~\ref{tab:impact-time-varying-HTE}: $0.020$ for transplant, $0.296$ for the age interaction, and $0.394$ for the year interaction. Transplant times follow a Cox model for time to transplant, fitted to the real data with age, year of acceptance, and prior surgery. Both models use the baseline hazards estimated from the real data. The simulated data sets closely resemble the real data: on average, they contain 75.1 deaths and 68.4 transplants, compared with 75 and 69 in the real data. We generate 500 data sets and apply four estimators: the two Cox models from the previous section and the two TV-CSL specifications used in Section~\ref{sec:copilot-application}, with nuisance models cross-fitted over five folds at the patient level. For each estimator, we report the bias and the coverage of nominal 95\% confidence intervals.

Table~\ref{tab:stanford-plasmode} reports the results. Ignoring transplant time is badly biased. On average, it underestimates the transplant effect by $1.549$ on the log-hazard scale, and its 95\% interval never contains the true value. This bias is about the same size as the difference between the two models in the real data ($-1.648$ versus $0.020$), so that difference is explained by immortal-time bias rather than by chance. The time-varying Cox model, which is correctly specified here, is nearly unbiased, with absolute bias at most $0.016$ and coverage between 0.95 and 0.96.

TV-CSL does not improve on the time-varying Cox model at this sample size. Its estimates vary about twice as much across data sets (empirical standard deviation 0.52--0.71, compared with 0.27--0.34) and are biased by 0.15--0.19. Most of this bias arises in the 32\% of data sets in which some training fold contains no prior-surgery deaths of one treatment status, which leaves the propensity poorly estimated in that fold; these datasets account for 0.13--0.16 of the bias, and the others for 0.02--0.03. Information-based intervals cover the true value 89\%--91\% of the time, but sandwich intervals cover it only 60\%--74\% of the time, because the sandwich standard error is roughly half the actual variability. These results agree with the main simulation. When baseline risk is simple and the sample is small, estimating the additional nuisance functions costs precision without reducing bias, and information-based intervals are more reliable than sandwich intervals. We therefore use the time-varying Cox model as the primary analysis for the Stanford data.

\input{tables/tab_stanford_plasmode}

\end{document}

%% file: tables/tab_copilot_ignore_time.tex

\begin{table}[ht]
\centering
\small
\setlength{\tabcolsep}{5pt}
\caption{Cox Estimates That Ignore and Include AI Assistant Initiation Time}
\label{tab:copilot-ignore-time}
\resizebox{\linewidth}{!}{%
\begin{tabular}{l|cc|cc}
\toprule
 & \multicolumn{2}{c|}{Ignore treatment time}
 & \multicolumn{2}{c}{Time-varying treatment} \\
\cmidrule(lr){2-3}\cmidrule(lr){4-5}
Effect & Hazard ratio (95\% CI) & $p$-value
       & Hazard ratio (95\% CI) & $p$-value \\
\midrule
Used AI Assistant (senior) & 2.11 (1.81, 2.47) & $<$0.001 & 1.63 (1.42, 1.88) & $<$0.001 \\
Used AI Assistant (junior) & 2.14 (1.59, 2.89) & $<$0.001 & 1.27 (1.08, 1.49) & 0.003 \\
Junior / senior ratio & 1.01 (0.72, 1.42) & 0.934 & 0.78 (0.64, 0.95) & 0.015 \\
\bottomrule
\end{tabular}
}
\par\smallskip
\begin{minipage}{\linewidth}
\footnotesize
Estimates are hazard ratios; values above 1 indicate a higher
instantaneous rate of becoming a regular Copilot user. The junior/senior ratio
and its confidence interval use the estimated covariance of the two group effects.
\end{minipage}
\end{table}

%% file: tables/tab_copilot_tvcsl.tex
\begin{table}[ht]
\centering
\small
\setlength{\tabcolsep}{4pt}
\caption{Comparison of Time-Varying Cox and TV-CSL Estimates}
\label{tab:copilot-tvcsl}
\resizebox{\linewidth}{!}{%
\begin{tabular}{lccc}
\toprule
 & Time-varying & \multicolumn{2}{c}{TV-CSL} \\
\cmidrule(lr){3-4}
Effect & Cox & Marginal propensity & Time-varying propensity \\
 & Hazard ratio (95\% CI) & Hazard ratio (95\% CI) & Hazard ratio (95\% CI) \\
\midrule
Used AI Assistant (senior) & 1.63 (1.42, 1.88) & 1.59 (1.35, 1.88) & 1.51 (1.28, 1.79) \\
Used AI Assistant (junior) & 1.27 (1.08, 1.49) & 1.18 (0.98, 1.42) & 1.23 (1.02, 1.48) \\
Junior / senior ratio & 0.78 (0.64, 0.95) & 0.74 (0.58, 0.94) & 0.81 (0.63, 1.04) \\
\bottomrule
\end{tabular}
}
\par\smallskip
\begin{minipage}{\linewidth}
\footnotesize
Estimates are hazard ratios; values above 1 indicate a higher
instantaneous rate of becoming a regular Copilot user. The junior/senior ratio
and its confidence interval use the estimated covariance of the two group effects.
TV-CSL (time-varying propensity) intervals use the cross-fitted sandwich variance estimator.
\end{minipage}
\end{table}

%% file: tables/tab_copilot_sensitivity.tex
\begin{table}[ht]
\centering
\small
\setlength{\tabcolsep}{4pt}
\caption{Sensitivity of AI Assistant Effect Estimates to Cohort and Tie Convention}
\label{tab:copilot-sensitivity}
\resizebox{\linewidth}{!}{%
\begin{tabular}{lcccc}
\toprule
 & \multicolumn{2}{c}{New users ($n = 2,026$)} & \multicolumn{2}{c}{All lawyers ($n = 3,322$)} \\
\cmidrule(lr){2-3}\cmidrule(lr){4-5}
Effect & $W(t) = 1(A < t)$$^\dagger$ & $W(t) = 1(A \le t)$ & $W(t) = 1(A < t)$ & $W(t) = 1(A \le t)$ \\
\midrule
\multicolumn{5}{l}{\textit{Cox, ignoring treatment time}} \\
\quad Used AI Assistant (senior) & 0.748 (0.591, 0.904) & 0.748 (0.591, 0.904) & 0.761 (0.616, 0.907) & 0.761 (0.616, 0.907) \\
\quad Used AI Assistant (junior) & 0.762 (0.461, 1.062) & 0.762 (0.461, 1.062) & 0.694 (0.413, 0.975) & 0.694 (0.413, 0.975) \\
\quad Junior $-$ senior difference & 0.014 (-0.322, 0.350) & 0.014 (-0.322, 0.350) & -0.067 (-0.380, 0.245) & -0.067 (-0.380, 0.245) \\
\addlinespace
\multicolumn{5}{l}{\textit{Time-varying Cox}} \\
\quad Used AI Assistant (senior) & 0.489 (0.348, 0.629) & 0.796 (0.660, 0.931) & 0.526 (0.416, 0.636) & 0.814 (0.704, 0.923) \\
\quad Used AI Assistant (junior) & 0.239 (0.081, 0.397) & 0.703 (0.540, 0.867) & 0.136 (0.015, 0.257) & 0.560 (0.429, 0.691) \\
\quad Junior $-$ senior difference & -0.250 (-0.452, -0.048) & -0.092 (-0.294, 0.110) & -0.390 (-0.540, -0.239) & -0.254 (-0.412, -0.096) \\
\addlinespace
\multicolumn{5}{l}{\textit{TV-CSL, marginal propensity}} \\
\quad Used AI Assistant (senior) & 0.466 (0.301, 0.631) & 0.718 (0.546, 0.890) & 0.507 (0.383, 0.631) & 0.764 (0.636, 0.892) \\
\quad Used AI Assistant (junior) & 0.165 (-0.019, 0.349) & 0.489 (0.303, 0.675) & 0.037 (-0.101, 0.174) & 0.423 (0.274, 0.571) \\
\quad Junior $-$ senior difference & -0.301 (-0.542, -0.060) & -0.229 (-0.476, 0.018) & -0.470 (-0.652, -0.289) & -0.342 (-0.535, -0.149) \\
\addlinespace
\multicolumn{5}{l}{\textit{TV-CSL, time-varying propensity}} \\
\quad Used AI Assistant (senior) & 0.415 (0.247, 0.582) & 0.658 (0.482, 0.833) & 0.439 (0.312, 0.566) & 0.693 (0.561, 0.824) \\
\quad Used AI Assistant (junior) & 0.205 (0.020, 0.390) & 0.576 (0.383, 0.768) & 0.046 (-0.095, 0.187) & 0.449 (0.292, 0.605) \\
\quad Junior $-$ senior difference & -0.210 (-0.455, 0.036) & -0.082 (-0.338, 0.173) & -0.394 (-0.582, -0.205) & -0.244 (-0.448, -0.041) \\
\bottomrule
\end{tabular}
}
\par\smallskip
\begin{minipage}{\linewidth}
\footnotesize
Estimates (95\% CI) on the log-hazard-ratio scale. $^\dagger$Primary analysis.
New users have no recorded AI Assistant use before time zero; the all-lawyers cohort
adds prior users and adjusts for prior use. Both cohorts exclude lawyers who became
regular users in their entry week. 
\end{minipage}
\end{table}

%% file: tables/tab_stanford_ignore_time.tex
\begin{table}[ht]
\centering
\caption{Cox Estimates That Ignore and Include Transplant Time}
\label{tab:impact-time-varying-HTE}
\resizebox{\linewidth}{!}{%
\begin{tabular}{l|cc|cc}
\toprule
 & \multicolumn{2}{c|}{Ignore transplant time}
 & \multicolumn{2}{c}{Include transplant time} \\
\cmidrule(lr){2-3}\cmidrule(lr){4-5}
Effect & Estimate (95\% CI) & $p$-value & Estimate (95\% CI) & $p$-value \\
\midrule
Transplant & -1.648 (-2.207, -1.089) & $<$0.001 & 0.020 (-0.616, 0.657) & 0.951 \\
Age $\times$ transplant & -0.228 (-0.782, 0.326) & 0.420 & 0.296 (-0.223, 0.816) & 0.263 \\
Year $\times$ transplant & 0.143 (-0.360, 0.645) & 0.578 & 0.394 (-0.116, 0.904) & 0.130 \\
\bottomrule
\end{tabular}
}
\par\smallskip
\begin{minipage}{\linewidth}
\footnotesize
Estimates are log hazard ratios for death; negative values indicate lower mortality
after transplant. Age and year of acceptance are standardized. All models adjust
for age, year of acceptance and prior surgery; prior surgery is not an effect modifier.
\end{minipage}
\end{table}

%% file: tables/tab_stanford_plasmode.tex
\begin{table}[ht]
\centering
\caption{Semi-Synthetic Bias and Coverage at the Stanford Sample Size}
\label{tab:stanford-plasmode}
\resizebox{\linewidth}{!}{%
\begin{tabular}{lcccccc}
\toprule
 & \multicolumn{2}{c}{Transplant ($0.02$)} & \multicolumn{2}{c}{Age $\times$ transplant ($0.30$)} & \multicolumn{2}{c}{Year $\times$ transplant ($0.39$)} \\
\cmidrule(lr){2-3}\cmidrule(lr){4-5}\cmidrule(lr){6-7}
Method & Bias & Coverage & Bias & Coverage & Bias & Coverage \\
\midrule
Cox, ignoring transplant time & -1.549 & 0.00 & -0.235 & 0.81 & -0.099 & 0.87 \\
Time-varying Cox & -0.016 & 0.96 & 0.009 & 0.96 & 0.008 & 0.95 \\
TV-CSL, marginal propensity & -0.171 & 0.91/0.74 & 0.159 & 0.90/0.64 & 0.193 & 0.90/0.71 \\
TV-CSL, time-varying propensity & -0.155 & 0.91/0.72 & 0.167 & 0.91/0.60 & 0.187 & 0.89/0.71 \\
\bottomrule
\end{tabular}
}
\par\smallskip
\begin{minipage}{\linewidth}
\footnotesize
Results over 500 replicates of $n = 103$. True values (in parentheses) are the
time-varying Cox estimates from the real data. Covariates and censoring dates are the
real ones; transplant and death times are redrawn from Cox models fitted to the real data.
Coverage is of nominal 95\% intervals; TV-CSL entries are information/sandwich.
\end{minipage}
\end{table}